\documentclass[twocolumn, superscriptaddress, eqseqnum]{revtex4-1}
\usepackage{amsmath,amsthm,amsfonts,amssymb}
\usepackage[normalem]{ulem}
\usepackage{graphicx}
\usepackage{psfrag}
\usepackage{multirow}
\usepackage{color}
\usepackage{natbib}
\usepackage{mciteplus}

\newcommand{\sigmabold}{\mbox{\boldmath{$\sigma$}}}
\newcommand{\ellbold}{\mbox{\boldmath{$\ell$}}}
\newcommand{\lambdabold}{\mbox{\boldmath{$\lambda$}}}
\renewcommand{\vec}[1]{{\underline{\mathbf #1}}}        
\newcommand{\uvec}[1]{\widehat{\mathbf #1}}
\newcommand{\ten}[1]{{\mathbf{#1}}}
\newcommand{\half}{\textstyle{\frac{1}{2}}}

\newcommand{\Id}{\ten{I}}

\newcommand{\tens}[1]{{\mathbf{#1}}}

\newcommand{\TT}{\tens{W}}
\begin{document}
 
\title{Tack energy and switchable adhesion of liquid crystal elastomers}

\author{D.~R.~Corbett}
\author{J.~M.~Adams} 
\affiliation{SEPnet and the Department of
  Physics, University of Surrey, Guildford, GU2 7HX, U. K. }

\date{\today}
\begin{abstract}
  The mechanical properties of liquid crystal elastomers (LCEs) make
  them suitable candidates for pressure-sensitive adhesives
  (PSAs). Using the nematic dumbbell constitutive model, and the block
  model of PSAs, we study their tack energy and the debonding process
  as could be measured experimentally in the probe-tack test. To
  investigate their performance as switchable PSAs we compare the tack
  energy for the director aligned parallel, and perpendicular to the
  substrate normal, and for the isotropic state. We find that the tack
  energy is larger in the parallel alignment than the isotropic case
  by over a factor of two. The tack energy for the perpendicular
  alignment can be $50\%$ less than the isotropic case. We propose a
  mechanism for reversibly switchable adhesion based on the
  reversibility of the isotropic to nematic transition.  Finally we
  consider the influence of several material parameters that could be
  used to tune the stress-strain response.
\end{abstract}
\maketitle
\section{Introduction}
\begin{figure}[!htb]
\begin{center}
\includegraphics[width = 0.48\textwidth]{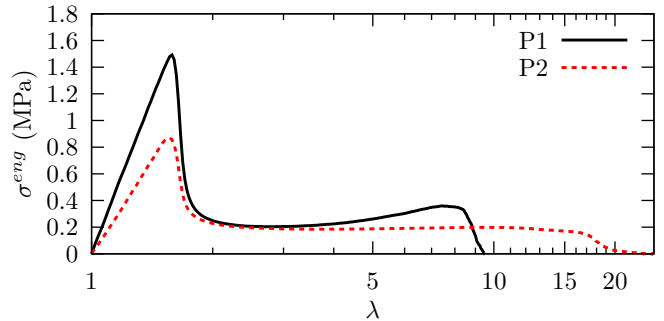}
\end{center}
\caption{The stress-deformation curve measured by a probe-tack test
  performed on two materials, P1: Poly(butyl acrylate) copolymer and
  P2: Poly(butyl acrylate) copolymer with CTA. Probe-tack measurements
  were performed at constant retraction velocity of $0.1$mm s$^{-1}$
  \cite{gurney2012}.}
\label{fig:gurney}
\end{figure}

Pressure sensitive adhesives (PSAs) are soft materials that adhere to
nearly any surface when low pressure is applied. Their effectiveness
can be described by the total energy required to separate the adhesive
from a surface, known as the \emph{tack energy}. Experimentally this
can be measured in the probe-tack test, where the force required to
remove a probe moving at constant velocity from an adhesive
film is measured and a stress-deformation curve is produced.  An
experimentally measured stress-deformation curve produced from a
probe-tack test on two materials is shown in fig.~\ref{fig:gurney}.
For small deformations the force rises rapidly with extension up to a
peak force. During this phase cavities form within the adhesive and
grow. Following the peak force there is a pronounced plateau in the
force which can be upward pointing (P1) if the material
strain-hardens. During this phase fibrillation occurs and the force is
dominated by the viscoelastic properties of the adhesive rather than
the cavities. Ultimately the plateau region ends when the material
detaches from the probe. Detachment can occur at the surfaces, which
is known as \emph{adhesive} failure and occurs for material P1 in
fig.~\ref{fig:gurney}. Alternatively detachment can happen in the
bulk, which is known as \emph{cohesive} failure and occurs for
material P2 in fig.~\ref{fig:gurney}. 

To achieve a high tack energy the PSA must have a low dynamic modulus
(typically $~0.1$ MPa at $1$ Hz) to make conformal contact with the
substrate~\cite{Dahlquist1969}. As the probe is retracted, the
adhesive film is drawn into fibrils. The adhesive must be sufficiently
soft so the fibrils do not detach at small strain, yet still require
some energy to deform \cite{deplacecarelli2009}. PSAs are typically
made from high molecular weight polymers that are lightly crosslinked
to form viscoelastic solids. Acrylic, styrenic, and siloxane based
polymers have been refined for use as PSAs \cite{lindner2006}. The
bulk rheological properties of a PSA are important in determining its
tack energy. The optimal stress-strain behaviour is initially strain
softening, to aid crack blunting, and eventually becomes strain
hardening to stiffen the fibrils, and increase the force required in
the latter stages of debonding \cite{deplacerabjohns2009}.

Gay \textit{et al.} have presented a successful theoretical
description of PSA debonding based on the growth of bubbles at the
substrate/adhesive interface \cite{gay1999}. These bubbles join,
forming larger bubbles which are then drawn into fibrils. The main
physical processes of homogeneous deformation, nucleation of cavities,
followed by fibril formation have been included in a simplified model
of the debonding process called the \emph{Block model}
\cite{yamaguchimorita2006,yamaguchidoi2006}. Here the adhesive layer
is divided up into $N$ equal rectangular blocks that can undergo a
combination of shear and stretching deformation, as well as slipping
on the substrate. The force-displacement curves of the block model
reproduce the characteristic features seen in probe-tack experiments.

There is considerable interest in being able to turn off the adhesive
properties of PSAs using external stimuli such as light, humidity, and
temperature. For example methacrylate-functionalized adhesives
containing a photoinitiator can show an almost complete loss of
adhesion when irradiated under a halogen lamp, as a result of
photo-initiated crosslinking raising the elastic
modulus~\cite{webster1999,Boyne:2001}. The switching here however is
only one-way, the adhesive cannot be returned to a tacky state.
Trenor \textit{et. al}~\cite{Trenor:2005} achieved two-way switching
by using courmarin-functionalized acrylate adhesives, whereby UV-A
radiation was used to switch off adhesion and UV-C radiation was used
to (partially) switch it back on. Altering ambient humidity has been
used to reversibly modify the surface composition of polymer blends
thus changing the adhesive strength~\cite{Diethert:2011}; note that
here the adhesion is modified by altering the substrate rather than
the adhesive itself. There are several other examples of switchable
adhesion/wetting based on polymer brush surfaces driven by an external
stimulus~\cite{Stuart:2010,Spina:2007}. Liquid Crystal Polymer (LCP)
based adhesives which undergo a reversible Smectic-Isotropic
transition have been shown to have a transition between a tacky and a
non-tacky regime as one cools to the smectic
phase~\cite{crevoisier1999}. The low temperature smectic phase is
harder, and less wetting than the isotropic phase.  Kamperman and
Synytska~\cite{Kamperman:2012} provide a review of two mechanisms to
achieve switchable adhesion; topography and chemical functionality.

In this paper we propose an application for weakly crosslinked LCPs,
called liquid crystal elastomers (LCEs), as anisotropic adhesives
whose tack energy can be switched depending on the orientation of the
director, and the degree of liquid crystalline order. The mechanism
here is based on the change in the bulk rheology of the adhesive.

LCEs are unique materials that couple liquid crystal mesogens to the
underlying polymer network. In the high temperature isotropic phase,
they behave much like conventional rubbers. When cooled into the
nematic phase the orientation of the director is crucial to
determining their mechanical behaviour. When stretched parallel to the
director they behave like uniaxial solids. On stretching perpendicular
to the director they exhibit (semi-)soft elasticity -- that is strain
increases with little increase in stress. They have a long plateau in
their stress-strain curve before undergoing strain hardening to values
of the modulus found in isotropic rubbers at low extension. The
nematic phase has been synthesized using both acrylate based
\cite{zubarev1996} and siloxane based polymer chemistry
\cite{kupfer1994}, though in practice the siloxane based LCEs have a
lower $T_g$ and more durable mechanical properties. Their mechanical
behaviour is well described by a phantom chain model of a network of
anisotropic Gaussian chains \cite{warnerterentjev2007}. This
description predicts the plateau in the stress-strain curve associated
with soft elasticity \cite{Olmsted1994}. However this equilibrium
statistical physics model does not describe the dynamics of the LCE
that are essential to model adhesive properties. We will show that the
nematic dumbbell model of Maffettone and Marrucci includes the
equilibrium behaviour of nematic elastomers, and also provides a
description of their dynamics \cite{1992maffettone}. By using this
constitutive model in the block model we describe the behaviour of LCE
based viscoelastic adhesives.

\section{Model System}

Here we consider an idealised version of the probe-tack test. A thin
adhesive layer is placed between two rigid substrates, which are then
pulled apart with a constant velocity. The force required to separate
the substrates is measured as a function of the displacement between
them. We will model the subsequent debonding process in two dimensions
by assuming that the adhesive and any deformational flow are confined
to the $xy$ plane. The debonding process involves several complex
phenomena, but the following dominant effects will be included in our
description; large strain deformation of the adhesive, cavity
expansion/contraction, slip of the adhesives at the substrate
interface. These effects have been incorporated in the Block model
developed by Yamaguchi \textit{et al.}~\cite{yamaguchimorita2006},
which gives a 2-dimensional description of isotropic adhesives at low
Reynolds number. We modify the Block model by utilising simpler shapes
and deformations for the blocks, and a material constitutive law that
describes LCEs and their associated director reorientation.

\subsection{Block Model}
\label{sec:blockmodel}

\begin{figure}[!htb]
\begin{center}
\includegraphics[width = 0.48\textwidth]{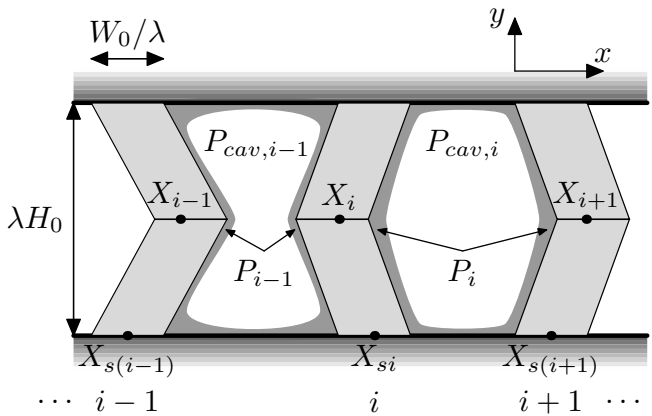}
\end{center}
\caption{The deformation of an individual block involves an area
  preserving elongation along the $y$ direction, and a piece-wise
  linear deformation in the $xy$ plane. The $x$ component of the
  centre of mass of the block is $X_{i}$ while the point of contact at
  the interface between the block and the substrate is $X_{si}$. 
}
\label{fig:blocks}
\end{figure}

The initial adhesive layer has height $H_{0}$ and length $L_{0}$ (we
assume the adhesive is thin, i.e. $L_{0}>>H_{0}$). This layer is
divided into $N$ rectangular blocks of equal width $W_{0}=L_0/N$,
which can undergo slip at the interface with the substrate. Each block
is assumed to undergo a stretching deformation along the
$y$-direction, no deformation in the $z$-direction, a corresponding
volume-conserving contraction in the $x$-direction, and a piece-wise
linear simple shear deformation in the $x$-direction as shown in
Fig.~\ref{fig:blocks}. The motion of the $i$th block is characterised
by three parameters: the elongation along the $y$-direction denoted
$\lambda$, the shear deformation denoted $\lambda_{xy,i}$ and the
$x$-coordinate of the centre of mass of the block denoted $X_{i}$.


On stretching the adhesive layer by a factor $\lambda$ in the $y$
direction, each block has a new height $H=\lambda H_{0}$ and a new
width $W=W_{0}/\lambda$.
The shear deformation of a block can be described by dividing it into
two sub-blocks of width $W$ and height $H/2$. The lower sub-block
shears by $\lambda_{xy}$ while the upper sub-block has the opposite
shear $-\lambda_{xy}$ as illustrated in Fig.~\ref{fig:blocks}. This
deformation is slightly different to that employed by
Yamaguchi~\textit{et al.} \cite{yamaguchimorita2006}. They allowed a
parabolic deformation in the $x$-direction, which implies a shear
which varies continuously with position along the $y$-axis.  We
anticipate a coupling between shears and the director in our nematic
adhesives, and modelling the nematic degrees of freedom is
substantially simplified if we assume that the sub-blocks have
constant shear and director. We will see later that this change simply
modifies the constant prefactor in the equation for the shear of the
block. The deformation gradient tensor $\lambdabold$ for each
sub-block can be written as
\begin{equation}
  \lambdabold=\left(\begin{array}{cc}\frac{1}{\lambda}& \pm\lambda_{xy}\\0& \lambda\end{array}\right),
\end{equation}
where the sign of the shear $\lambda_{xy}$ depends upon which
sub-block is being described.

Let $(X_{si},0)$ be the mid-point of the contact between the block and
the substrate in the deformed state, and $(x_{si},0)$ the equivalent
point in the undeformed state. Then transforming between the
undeformed state $(x,y)$ and deformed state $(X, Y)$ using
$(dX,dY)=\tens{\lambdabold}\cdot(dx,dy)$ and integrating we obtain
\begin{eqnarray*}
 X_{i}(x,y)&=&\left\{\begin{array}{cc}X_{si}+\frac{(x-x_{si})}{\lambda}+\lambda_{xy,i}y&y<H_{0}/2\\
   X_{si}+\frac{(x-x_{si})}{\lambda}+\lambda_{xy,i}(H_{0}-y)&y\geq H_{0}/2\end{array}\right.\\
 Y(x,y)&=&\lambda y.
\end{eqnarray*}
Hence the $x$-coordinate of the centre of mass of the $i$th block, is
given by
\begin{eqnarray}
X_{i}&=&\int_{y=0}^{y=H_{0}}\int_{x=x_{si}-W_{0}/2}^{x=x_{si}+W_{0}/2}X(x,y)dxdy
/(H_{0}W_{0})\nonumber\\
&=&X_{si}+\lambda_{xy,i}\frac{H_{0}}{4}.
\end{eqnarray}
Finally, we will require the elements of the velocity gradient tensor
in the deformed configuration $K_{ij}=\nabla_j v_i$. In terms of the
deformation gradient tensor $\lambdabold$ this is given by
$\tens{K}=\tens{\dot{\lambdabold}}\cdot\lambdabold^{-1}$:
\begin{equation}
 \tens{K}=\left(
\begin{array}{cc}-\frac{\dot{\lambda}}{\lambda}& \;\;\;\mp\left[\frac{\dot{\lambda}}{\lambda}+\frac{\dot{\lambda}_{xy}}{\lambda_{xy}}\right]\frac{\lambda_{xy}}{\lambda}\\0& \frac{\dot{\lambda}}{\lambda}\end{array}
\right).\label{eq:Kelements}
\end{equation}

\subsection{Cavity Expansion}

\begin{figure}[!htb]
\begin{center}
\includegraphics[width = 0.48\textwidth]{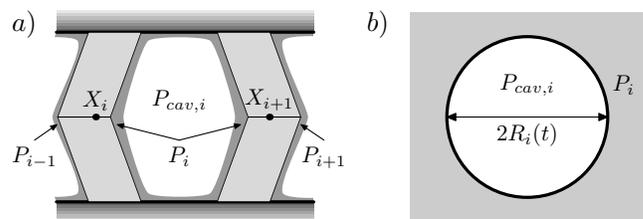}
\end{center}
\caption{(a) If the difference $(X_{i+1}-X_{i})$ is greater than
  $W_{0}/\lambda$ we say a cavity has formed between the blocks as
  illustrated. The pressure $P_{i}$ on the faces of blocks $i$ and
  $i+1$ is required to calculate the debonding force. (b) We use a
  simplified model of the cavity, assuming it to be a spherical bubble
  of radius $R_{i}$ placed in an infinite Newtonian fluid with
  viscosity $\eta=G\tau$. Assuming creep flow and applying force
  balance at the interface between the bubble and the interface leads
  to a value for $P_{i}$, the pressure on the fluid side of the bubble
  interface.}
\label{fig:bubble}
\end{figure}

As the separation between substrates $\lambda H_{0}$ increases, there
is a corresponding reduction in block width $W_{0}/\lambda$. If the
displacement between the centres of the blocks $(X_{i+1}-X_{i})$ is
not equal to their width, a gap is created between the blocks labelled
$i+1$ and $i$. We treat this gap as a circular cavity with radius
$R_{i}$ having the same area as the void between the blocks, see
Fig.~\ref{fig:bubble}. The instantaneous radius of the circular
bubble is given by
\begin{equation}
\pi R_{i}^{2}=H_{0}\lambda\left(X_{i+1}-X_{i}-\frac{W_{0}}{\lambda}\right).\label{eq:geompos}
\end{equation}

The Rayleigh-Plesset (RP) equation \cite{rayleigh1917,plesset1977}
describes the dynamics of a spherical bubble placed in a Newtonian
fluid with viscosity $\eta$. If $P_{i}$ is the pressure in the fluid
at the bubble interface, $P_{cav,i}$ the pressure in the cavity at the
fluid interface and $\gamma$ the surface tension, then the RP equation
is
\begin{equation}
  \frac{dR_{i}}{dt}=\frac{R_{i}}{2\eta}\left(P_{cav,i}-P_{i}-\frac{\gamma}{R_{i}}\right).\label{eq:cavdyn}
\end{equation}
If the cavities do not contain air then $P_{cav,i}$ is simply zero. If
the cavity includes air, initially at atmospheric pressure
$P_\textrm{atm}$ in the undeformed state then $P_{cav,i}$ is related
to the area change in the cavity via
\begin{equation}
P_{cav,i}(t)=P_\textrm{atm}\left(\frac{R_{i}(0)}{R_{i}(t)}\right)^{2}.
\end{equation}

This is a highly simplified model of the growth/contraction of the
cavities, it ignores both the elasticity and the anisotropy of the
viscoelastic adhesive surrounding the cavity.  Given the anisotropy of
the adhesive it is unlikely bubbles would grow in a uniform
circular/spherical way.  Nevertheless the model has the correct
qualitative features for any model of cavity dynamics and we adopt it
for simplicity.
\subsection{Block positions}

Equation~(\ref{eq:geompos}) gives the radii of the $N-1$ cavities in
terms of the positions of $N$ blocks. To invert this equation and
hence determine the positions of the $N$ blocks we need an additional
constraint. The applied external stretching force is in the $y$
direction, there is no external force in the $x$ direction. The $x$
coordinate of the centre of mass of the $N$ blocks thus does not
change with time
\begin{equation}
\sum_{i=1}^{N}X_{i}(t)=\sum_{i=1}^{N}X_{i}(t=0)=0.\label{eq:xcompos}
\end{equation}
Therefore if we know the cavity radii, we can use
equations~(\ref{eq:geompos}) and(~\ref{eq:xcompos}) to solve for the
positions of all $N$ blocks.

\subsection{Stress Tensor}

The total stress tensor within each block is made up of two
components, the polymer stress arising from forces transmitted by the
polymer chains within the adhesive denoted by $\tens{\Sigma}$, and the
isotropic pressure term $-p\Id$. The total stress tensor
$\sigmabold$ is produced by adding these two components
\begin{equation}
\sigmabold=-p\Id+\tens{\Sigma}.
\label{eqn:sigma}
\end{equation}
The constitutive equation obeyed by $\ten{\Sigma}$ will be discussed
in \S \ref{sec:constit}.  We assume that the components of the
polymer stress tensor are homogenous within each block. This is not so
for the pressure contribution. The $i$th block has two free
surfaces. At these surfaces the $xx$ component of the total stress
tensor is $\sigma_{xx}=-P_{i}$ and $\sigma_{xx}=-P_{i+1}$. To
accommodate this change in the total stress it is clear the term
$-p\Id$ must vary across the block. We will assume this variation is
linear.

\subsection{Slippage at the interface}
We denote by $\sigma_{si}$ the shear stress at the interface between
the $i$th block and the substrate. We assume a simple linear
relationship between the position of the interface between the block
and the substrate $X_{si}$ and the shear stress $\sigma_{si}$ at the
same place
\begin{equation}
\mu \frac{d}{dt}X_{si}=\sigma_{si}.\label{eq:slip}
\end{equation}
Recall $X_{si}=X_{i}-\lambda_{xy,i}H_{0}/4$, thus having previously
determined the positions of the blocks $X_{i}$ we can rewrite this as
an equation for the rate of change of the shear strain
$\lambda_{xy,i}$.

\subsection{Force Balance}

The difference in the pressure on either side of the block
$(P_{i+1}-P_{i})$ is balanced by the shear stress, leading to a force
balance equation
\begin{equation}
(P_{i+1}-P_{i})H_{0}\lambda=-2\sigma_{si}\frac{W_{0}}{\lambda}.\label{eq:pressuredif}
\end{equation}

Once the shear stress of each block has been calculated, we can use this equation, coupled with
the boundary condition $P_{1}=P_\textrm{atm}$ to calculate the pressure within each block.

\subsection{Debonding Force}

To determine the debonding force $F_{y}$ (the total force acting on
the substrate) we require the pressure, and the $xx$ and $yy$
components of the polymer stress tensor $\tens{\Sigma}$.  The
$\sigma_{xx}$ component at the left and right edges of the $i$th
block can be used to calculate an approximate value of the average
pressure in the block $p$ using the assumption of a linear variation
in $p$ through the block
\begin{eqnarray}
\sigma_{xx,i}\big|_{x_{si}-W_0/2} &=& \Sigma_{xx,i} - p\big|_{x_{si}-W_0/2} = -P_{i-1}\\
\sigma_{xx,i}\big|_{x_{si}+W_0/2} &=& \Sigma_{xx,i} - p\big|_{x_{si}+W_0/2} = -P_{i}\\
\Rightarrow p &=& p\big|_{x_{si}-W_0/2}+p\big|_{x_{si}+W_0/2}\nonumber \\
&\approx& \Sigma_{xx,i} + \frac{P_{i}+P_{i-1}}{2}
\label{eqn:blockp}
\end{eqnarray}
The $\sigma_{yy,i}$ component from the $i$th block is thus
\begin{equation}
\sigma_{yy,i} = \Sigma_{yy,i} - \Sigma_{xx,i}- \frac{P_{i}+P_{i-1}}{2}.
\end{equation}
This result contains the normal stress difference of $\ten{\Sigma}$,
hence the addition of an isotropic pressure term to the constitutive
model will be absorbed into the pressure $p$ of
Eq.~(\ref{eqn:blockp}). The total force $F_y$ can be calculated by
summing the force due to atmospheric pressure on the substrates, and that
due to each block in the adhesive
\begin{equation}
F_{y}=\frac{A}{N}\sum_{i=1}^{N}\left(\Sigma_{yy,i}-\Sigma_{xx,i}-\frac{P_{i}+P_{i+1}}{2}\right)+P_\textrm{atm}A, \label{eqn:fy}
\end{equation}
where $A$ is the initial contact area. Note Yamaguchi \textit{et al.}
did not use a normal stress difference in \cite{yamaguchimorita2006},
however it makes little difference to the results for the parameters
and strain ranges used there. Dividing Eq.~(\ref{eqn:fy}) through by
$A$ we obtain the engineering stress in debonding
\begin{equation}
\sigma_{yy}^\textrm{eng}=\frac{F_{y}}{A}=\overline{(\Sigma_{yy}-\Sigma_{xx})}+(P_\textrm{atm}-\overline{P}),\label{eq:debond}
\end{equation}
where $\overline{T}$ represents an average of quantity $T$ over all
blocks. To calculate the tack energy $\mathcal{E}_{T}$ we integrate the
debonding force over the distance moved by the upper substrate
\begin{equation}
\mathcal{E}_{T}=\int_{y'=H_{0}}^{y'=y} F_{y'} dy' = AH_{0}\int_{\lambda'=1}^{\lambda'=\lambda}\sigma(\lambda')d\lambda'.
\label{eq:tack}
\end{equation}

\section{Constitutive Relationship}
\label{sec:constit}

To determine the tack energy of the PSA we require the rheological
properties of the adhesive layer in \S \ref{sec:blockmodel}. As we are
modelling a LCE based PSA, we will use the nematic dumbbell model of
Maffettone and Marrucci~\cite{1992maffettone}. We consider a polymer
chain made up of $N$ freely hinged nematogenic rods with individual
length $b$. The end-to-end vector for the chain is $\vec{R}$. Assuming
Gaussian fluctuations the quantity of interest is the scaled second
moment of the end-to-end vector, that is
$\TT=3\left<\vec{R}\,\vec{R}\right>/Nb^{2}$. The dynamics of $\TT$ are
governed by
\begin{eqnarray}
\stackrel{\nabla}{\TT}&=&\frac{1}{\tau}\Id-\frac{1}{2\tau}\left(\ellbold^{-1}\cdot\TT+\TT\cdot\ellbold^{-1}\right),\label{eq:scaleddynamicsRR}\\
\tens{\Sigma}&=&G\ellbold^{-1}\cdot\TT,\label{eq:scaledpolymerstress}
\end{eqnarray}
where
$\stackrel{\nabla}\TT=d\TT/dt-\tens{K}\cdot\TT-\TT\cdot\tens{K}^{T}$
is the upper convected Maxwell derivative (UCMD) and $\tens{K}$ is the
velocity gradient tensor in Eq.~(\ref{eq:Kelements}). The inverse
chain shape tensor $\ellbold^{-1}$ is given by
\begin{equation}
\ellbold^{-1}=\frac{1}{(1-S)}\left(\Id-\frac{3S}{1+2S}\uvec{n}\,\uvec{n}\right),
\end{equation}
in which the nematic director $\uvec{n}$ describes the average
orientation of the nematogenic units and the order parameter $S$ which
describes the degree of alignment along the director ($S=1$
corresponds to perfect ordering, while $S=0$ corresponds to an
isotropic phase). The polymer stress $\tens{\Sigma}$ is that arising
from the polymer chains within the material. The time-scale $\tau$ is
related to the diffusivity of the chain ends $D$ via $\tau = Nb^{2}/6D$,
while the modulus 
\begin{equation}
G=ck_{B}T,
\label{eqn:modulus}
\end{equation}
where $c$ is the number of chains per unit volume.

This Gaussian model does not include effects of finite chain
extensibility, or entanglements. However Gaussian models have been
shown to provide a good description of the mechanical behaviour of
liquid crystalline elastomers \cite{warnerterentjev2007}.

\subsection{Isotropic Limit}
Yamaguchi \textit{et al.} first described the block model using the
constitutive equation of an isotropic Maxwell fluid for the polymer
stress. Here we demonstrate the isotropic limit of
Eqs.~(\ref{eq:scaleddynamicsRR}) and~(\ref{eq:scaledpolymerstress})
produce the same constitutive equation.

To obtain the isotropic limit, we set $S=0$ in Eq.~(\ref{eq:scaleddynamicsRR}) and~(\ref{eq:scaledpolymerstress}). This gives
\begin{eqnarray}
\stackrel{\nabla}{\TT}&=&\frac{1}{\tau}\Id-\frac{1}{\tau}\TT,\label{eq:isodynamicsRR}\\
\tens{\Sigma}&=&G\TT.\label{eq:isopolymerstress}
\end{eqnarray}
Substituting for $\TT$ in equation~(\ref{eq:isodynamicsRR}) in terms of $\tens{\Sigma}$
\begin{equation}
\stackrel{\nabla}{\tens{\Sigma}}=\frac{G}{\tau}\Id-\frac{1}{\tau}\tens{\Sigma}.
\end{equation}
Any isotropic stress can be added to the definition of $\tens{\Sigma}$,
as it can be absorbed into the pressure term in Eq.~(\ref{eqn:sigma}).
It is convenient to subtract $G\Id$ from the stress
\mbox{$\tens{\Sigma}^\prime=\tens{\Sigma}-G\Id$}. This subtraction is
useful since $\tens{\Sigma}^\prime$ is zero in the undeformed state.
\begin{equation}
\stackrel{\nabla}{\tens{\Sigma}^\prime}+\frac{\tens{\Sigma}^\prime}{\tau}=2G\tens{D},
\end{equation}
where $\tens{D}=(\tens{K}+\tens{K}^{T})/2$ is the symmetric part of
the velocity gradient tensor.  This is identical to the constitutive
relationship used by Yamaguchi \textit{et al.}
\cite{yamaguchimorita2006}.

\subsection{Director dynamics}

To complete our description of the dynamics of the polymer stress we
must also describe the behaviour of the nematic director with time. In
general this is a complicated task involving the coupling of the
director to the flow field. Maffetone and Marrucci
\cite{1992maffettone} identify two simplified regimes
\begin{enumerate}
\item \textit{Weak external field.} In this case the polymer stress
  tensor is required to be symmetric. To achieve this in
  Eq.~(\ref{eq:scaledpolymerstress}) the director $\uvec{n}$ must be
  one of the eigenvectors of $\TT$. Since in static equilibrium we
  require the polymer stress to be isotropic, $\TT_{eq}$ must be equal
  to $\ellbold$, from which we can identify $\uvec{n}$ as the
  eigenvector corresponding to the largest eigenvalue of $\TT$.

\item \textit{Strong external field.} In this case the torques arising
  from coupling to the flow field are insufficient to move $\uvec{n}$
  from the direction imposed by the external field. The polymer stress
  will not be symmetric in this case.

\end{enumerate}

We model the first of these limits, where the director responds much
faster than the polymers. We leave the more complicated task of
generalised director dynamics to future work.

\subsection{Quasi-Static Limit}

There are consistent theoretical descriptions of nematic elastomers
derived phenomenologically from continuum mechanics
\cite{PhysRevE.66.011702}, and from a microscopic equilibrium
statistical physics model \cite{warnerterentjev2007}. These models
both contain the Goldstone modes predicted in Nematic elastomers using
symmetry arguments \cite{PhysRevLett.63.1082}, known as \emph{soft
  modes}. The microscopic model produces the following trace formula
for the free energy density $F$
\begin{equation}
F = \half G \textrm{Tr} \left[\lambdabold\cdot \ellbold_0 \cdot \lambdabold^T\cdot \ellbold^{-1} \right].
\label{eqn:trace}
\end{equation}
where $G$ is the shear modulus of the rubber defined in
Eq.~(\ref{eqn:modulus}), $\ellbold_{0}$ is the initial chain shape
distribution, and it is assumed that $\textrm{det}[ \lambdabold] =
1$. The soft modes permitted by this free energy have the following
explicit form
\begin{equation}
\lambdabold=\ellbold^{\half}\cdot\tens{Q}\cdot\ellbold_{0}^{-\half},
\label{eqn:lambdasoft}
\end{equation}
where $\tens{Q}$ is an orthogonal tensor. They arise because states
which are related to each other by a simple rotation of the chain
shape distribution tensor $\ellbold$ have the same energy. The true
stress can be derived from Eq.~(\ref{eqn:trace}) by differentiating
$F$ with respect to $\lambdabold$, then post-multiplying by
$\lambdabold^T$, producing a polymer stress component
\begin{equation}
  \tens{\Sigma} = G \ellbold^{-1}\cdot \lambdabold\cdot \ellbold_0 \cdot \lambdabold^T.
\label{eqn:stresssm}
\end{equation}
Eqs.~(\ref{eq:scaleddynamicsRR}) and~(\ref{eq:scaledpolymerstress})
also permit soft mode solutions. Motivated by Eq.~(\ref{eqn:stresssm})
we substitute $\TT=\lambdabold\cdot\ellbold_{0}\cdot\lambdabold^{T}$,
into Eq.~(\ref{eq:scaleddynamicsRR}) and observe that the UCMD is
identically zero, leaving 
\begin{eqnarray}
  0&=&\frac{\Id}{\tau}-\frac{(\ellbold^{-1}\cdot\lambdabold\cdot\ellbold_{0}\cdot\lambdabold^{T}+\lambdabold\cdot\ellbold_{0}\cdot\lambdabold^{T}\cdot\ellbold^{-1})}{2 \tau}.
\end{eqnarray}
This equation holds provided that $\lambdabold$ obeys
Eq.~(\ref{eqn:lambdasoft}). The stress tensor associated with these
modes is simply $\tens{\Sigma}=G\Id$. The normal stress difference
$\Sigma_{yy}-\Sigma_{xx}=0$, and thus these modes would not contribute
directly to the debonding force of Eq.~(\ref{eq:debond}) (they
contribute indirectly in so much as the pressure difference across the
block is determined by the shear stress associated with the mode). In
typical experiments on nematic elastomers stretching perpendicular to
the nematic director produces a soft response as the director rotates
towards the stretch direction. Stretching parallel to the director
produces a hard elastic response. We thus have reason to suspect that
the debonding force could be substantially different for parallel and
perpendicular geometries.

\section{Numerical Method}

To solve our set of differential equations we implement a mixed
implicit/explicit finite difference scheme to step forwards in time.
To improve computational accuracy we reduce the number of free
parameters by combining those with equivalent effects. The numerical
results are in agreement with the analytic single block results and
the semi-analytic two block case with $\tau \rightarrow \infty$. They
are consistent with the isotropic results of Yamaguchi \textit{et al.}
\cite{yamaguchimorita2006}.

\subsection{Initial Condition}
The initial condition has $N$ blocks all of which have $\lambda=1$ and
$\lambda_{xy,i}=0$. There are $N-1$ cavities with initial radius
$R_{i}=R_{0}\exp(\chi_{i})$ where $\chi_{i}$ is assigned from a
normal distribution with $\left<\chi\right>=0$ and
$\left<\chi^{2}\right>=1$ and $R_{0}$ is a characteristic size. The
initial positions \mbox{$X_{i}(t=0)$} can then be determined by solving
Eqs.~(\ref{eq:geompos}) and~(\ref{eq:xcompos}). The pressures at the
block interfaces are initially set to atmospheric pressure, thus
$P_{i}=P_\textrm{atm}$. We assume that the cavities form at the
substrate/adhesive interface and are initially filled with air, and
thus we also set the cavity pressure equal to atmospheric pressure,
$P_{cav,i}=P_\textrm{atm}$~\cite{ISI:000232026600001}.  Experimentally, the cavities are sometimes found to be filled with vacuum even if the cavities formed at the substrate/adhesive interface~\cite{glassmaker2008}.  Yamaguchi \textit{et. al} found the presence or absence of air within the cavities made very little difference to the resulting tack curves~\cite{yamaguchimorita2006}.  We divide the time $t$ into units $\delta t$ and advance the equation set using a mixed explicit/implicit scheme.  We introduce a superscript to quantities to indicate at which time-step they are evaluated, i.e. $\lambda^{(n)}$ is evaluated at the $n$th time step when $t=n\delta t$.

\subsection{Time Stepping}

\begin{enumerate}
\item We increment the strain explicitly 
\mbox{  $\lambda^{(n+1)}=\lambda^{(n)}+\dot{\lambda}\delta t$}.
\item We update the cavity radius semi-implicitly 
  \begin{equation}
R_{i}^{(n+1)}=\frac{2\eta R_{i}^{(n)}-\gamma \delta t}{2\eta+\delta t(P_{i}^{(n)}-P_{cav,i}^{(n)})}.\nonumber
\end{equation}
\item Using the new values of the cavity radii we obtain the cavity
  pressure
\begin{equation}
P_{cav,i}^{(n+1)}=P_{cav,i}^{(n)}\left(\frac{R_{i}^{(n)}}{R_{i}^{(n+1)}}\right)^{2}.\nonumber
\end{equation}
\item With the new values for the strain $\lambda^{(n+1)}$ and cavity radii $\{R_{i}^{(n+1)}\}$ we solve Eqs.~(\ref{eq:geompos}) and~(\ref{eq:xcompos}) for the new positions of the blocks $\{X_{i}^{(n+1)}\}$.

\item Using $X_{si}=X_{i}-\lambda_{xy,i}H_{0}/4$ we update
  $\lambda_{xy,i}$ from Eq.~(\ref{eq:slip}) 
\begin{equation}
\lambda_{xy,i}^{(n+1)}=\lambda_{xy,i}^{(n)}+\frac{4}{H_{0}}\left[(X_{i}^{(n+1)}-X_{i}^{(n)})-\frac{\sigma_{si}^{(n)}}{\mu}\delta{t}\right].\nonumber
\end{equation}
This equation replaces the curvature of the blocks used by Yamaguchi
\textit{et al.}. It describes the same physical process of block
shear, though it differs by a geometrical factor.

\item With the shear strains $\{\lambda_{xy}^{(n+1)}\}$ and
  $\lambda^{(n+1)}$ we calculate the elements of the deformation
  gradient tensor
\begin{eqnarray*}
K_{yy\hphantom{,i}}&=&\frac{\dot{\lambda}}{\lambda^{(n+1)}},\\
K_{xx\hphantom{,i}}&=&-K_{yy},\\
K_{xy,i}&=&-\left[\frac{K_{yy}}{\lambda^{(n+1)}}+\frac{(\lambda_{xy,i}^{(n+1)}-\lambda_{xy,i}^{(n)})}{\lambda^{(n+1)}\delta t}\right],\\
K_{yx\hphantom{,i}}&=&0.
\end{eqnarray*}
\item We now update elements of $\TT$ using a mixed explicit/implicit scheme
\begin{eqnarray*}
W_{yy}^{(n+1)}&=&\frac{\left[W_{yy}^{(n)}+\frac{\delta t}{3}-\alpha\frac{\delta t}{1-S} n_{x}^{(n)}n_{y}^{(n)}W_{xy}^{(n)}\right]}{1-2K_{yy}\delta t+\frac{\delta t}{1-S}\left[1+\alpha n_{y}^{(n)}n_{y}^{(n)}\right]},\\
W_{xy}^{(n+1)}&=&\frac{1}{1+\frac{\delta t}{2(1-S)}\left[1+\left(\frac{1}{r}+1\right)n_{x}^{(n)}n_{x}^{(n)}\right]}\\
\times &&\left(W_{xy}^{(n)}-\alpha\delta t \frac{n_{x}^\textrm{(n)}n_{y}^\textrm{(n)}(W_{xx}^{(n)}+W_{yy}^{(n+1)})}{2(1-S)}\right.\\
&&\left.+\delta t K_{xy}W_{yy}^{(n+1)}\right)\\
W_{xx}^{(n+1)}&=&\frac{1}{1+2K_{yy}\delta t+\frac{\delta t}{1-S}\left[1+\alpha n_{x}^{(n)}n_{x}^{(n)}\right]}\\
\times &&\left(W_{xx}^{(n)}+\frac{\delta t}{3}-\alpha\delta t \frac{n_{x}^{(n)}n_{y}^{(n)}W_{xy}^{(n+1)}}{1-S}\right.\\
&&\left.+2\delta tK_{xy}W_{xy}^{(n+1)}\right),
\end{eqnarray*}
where $\alpha = -\frac{3S}{2S + 1}$.
\item Determine the director by finding the eigenvector of
  $\TT^{(n+1)}$ corresponding to the largest eigenvalue.
\item Calculate the updated polymer stress $\tens{\Sigma}^{(n+1)}$
  from Eq.~(\ref{eq:scaledpolymerstress}).
\item Calculate the new pressure field $\{P_{i}^{(n+1)}\}$ from
  Eq.~(\ref{eq:pressuredif}) with the boundary condition
  $P_{1}=P_\textrm{atm}$.
\item Repeat to advance to the next time step.
\end{enumerate}

\section{Results and Discussion}

A typical set of simulation parameters are shown in
Table \ref{table:one}.  Most of the values are those used by Yamaguchi
and Doi~\cite{yamaguchimorita2006}.  According to the Maier-Saupe
theory~\cite{Maier:59} of the isotropic-nematic phase transition, the
order parameter at the first order jump to the nematic phase is
$S=0.42$.  The orientational order parameter associated with the
polymer backbone is usually a fraction of this bare nematic order
parameter, which motivates our choice of $S=0.3$.  We should note
however that much higher values of the order parameter have been
reported for the backbone order parameter in LCEs, particularly for
main-chain systems where $S=0.9$ has been
observed~\cite{Tajbakhsh:01}.  The value of $\tau$ adopted here is
lower than that used by Yamaguchi \textit{et. al.} but is consistent
with creep flow measurements on adhesives manufactured from acrylic polymers \cite{Degrandi:09}.

\begin{figure*}[!htb]
\includegraphics[width=\textwidth]{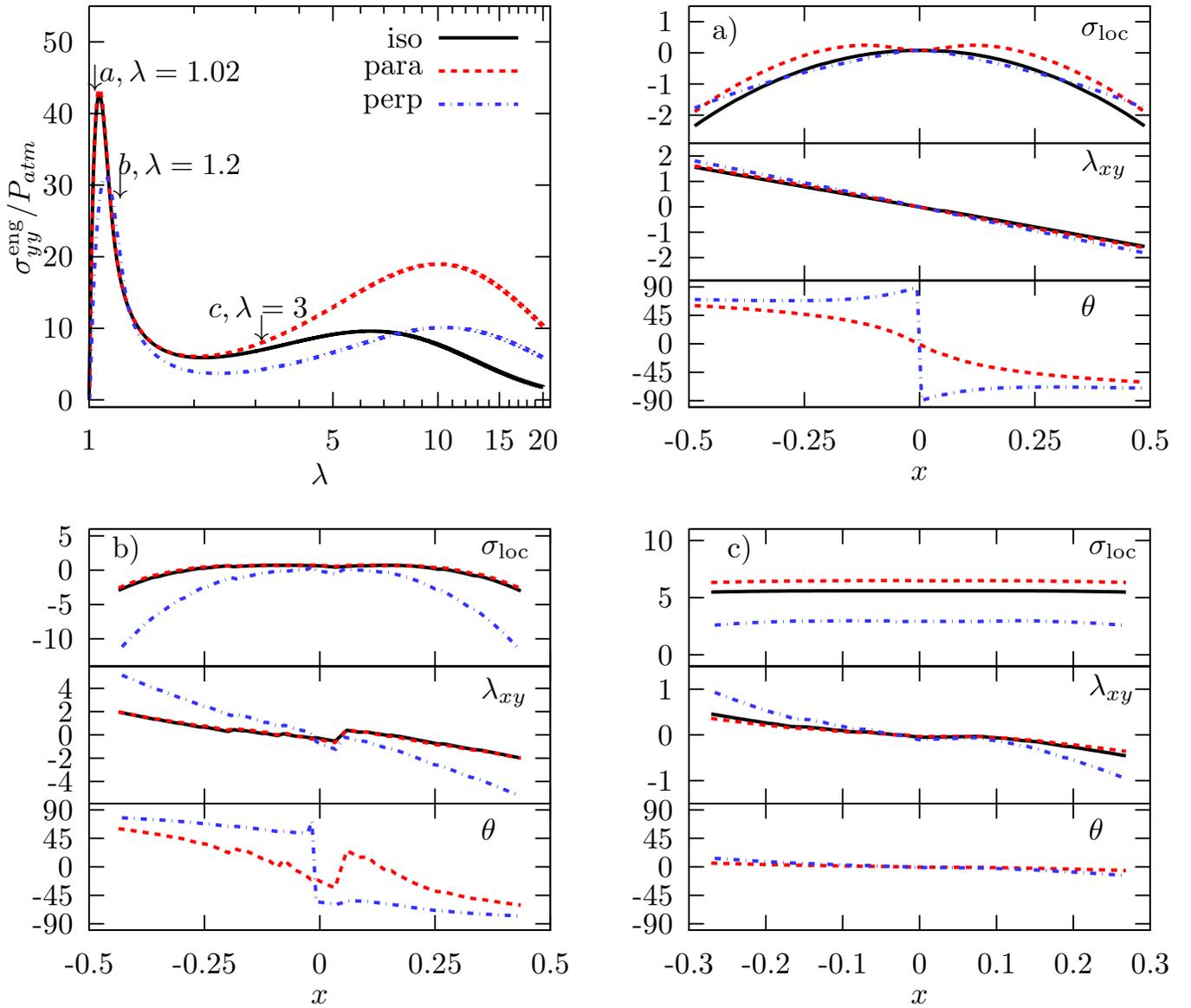}
\caption{The debonding stress $\sigma_{yy}^\textrm{eng}$ as a function of the
  deformation $\lambda$ for the parallel, isotropic and perpendicular
  arrangements. The debonding stress $\sigma_{loc}$ in units of
  $P_\textrm{atm}$, the shear $\lambda_{xy}$ and the angle the
  director makes with the $y$-axis as functions of position within the
  adhesive $x$ are shown in (a) for $\lambda=1.02$, (b) for
  $\lambda=1.2$ and (c) for $\lambda=3$.}
\label{fig:results}
\end{figure*}

\begin{table}[!htb]
\begin{tabular}{ll}
\hline
Parameter (symbol) & Value\\ 
\hline
Atmospheric Pressure ($P_\textrm{atm}$) & $10^{5}$ Pa\\
Shear Modulus ($G$) & $10^{5}$ Pa\\
Relaxation Time ($\tau$) & $30$ s\\
Viscosity ($\eta = G\tau$) & $3\times 10^{6}$ Pa s\\
Strain Rate ($\dot{\lambda}$) & $0.1$ s$^{-1}$\\
Surface Tension ($\gamma$) & $3\times 10^{-2}$ J m$^{-2}$\\
Friction Coefficient ($\mu$) & $2\times 10^{9}$ Pa s m$^{-1}$\\
Typical Cavity Radius ($R_{0}$) & $10^{-6}$ m\\
Initial Height ($H_{0}$) & $10^{-4}$ m\\
Length ($L_{0}$) & $5\times 10^{-3}$ m\\
Order Parameter ($S$) & 0.3 (0 if isotropic)\\
Number of Blocks ($N$) & 100\\
\hline
\end{tabular}
\caption{Model Parameters}
\label{table:one}
\end{table}
In Fig.~\ref{fig:results}(a) we show how the debonding stress
$\sigma_{yy}^\textrm{eng}$ varies with the deformation $\lambda$
(notice the logarithmic scale on the $x$-axis) for three cases; (i)
the adhesive is isotropic, (ii) the adhesive is nematic with the
director initially parallel to the stretch direction and (iii) the
adhesive is nematic with the initial director perpendicular to the
stretch direction. Each case has the same initial random seed for
cavity formation.
The three cases show the typical behaviour of a PSA.  We show the
behaviour up to a very large deformation of $\lambda = 20$. Whilst
such large deformations can be achieved experimentally using soft,
weakly crosslinked polymers \cite{deplacerabjohns2009} the Gaussian
model used here does not provide a good description of these large
deformations, or the detachment of the adhesive. However, it does
provide a qualitative model of the subsequent behaviour that may be
achieved at lower strains in more precise models.  We can identify $4$
separate behaviours within the plot:
\begin{enumerate}
\item A steep initial rise in the debonding stress until $\lambda\sim 1.05$.
\item A decrease in the debonding stress until $\lambda \sim 2$.
\item An increase in the debonding stress until $\lambda \sim 5$ for
  the isotropic case and $\lambda\sim 10$ for nematic cases.
\item A decrease in the debonding stress for larger strains.
\end{enumerate}

We explain the main processes occurring during each of these stages.
Fig.~\ref{fig:results}(a) shows the local debonding stress
\begin{equation}
  \sigma_{loc}=\left[\Sigma_{yy,i}-\Sigma_{xx,i}+P_\textrm{atm}-
    \frac{P_{i}+P_{i+1}}{2}\right]/P_\textrm{atm},
\end{equation}
the shear strain $\lambda_{xy}$ and the angle $\theta$ that the
director makes with the $y$-axis as a function of position in the
adhesive for $\lambda=1.02$.  The local debonding stress is small near
the edges ($x\sim \pm 0.5$) of the adhesive and rises rapidly towards
the centre ($x\sim 0$).  Investigating separately the pressure and
polymer stress contributions to the local debonding stress reveals the
pressure to be the predominant cause of the initial rise in the
debonding stress.  As each block expands along the $y$-direction it
must also contract along the $x$-direction.  This contraction results
in large shear strains and concomitantly large shear stresses. These
shear stresses are balanced by the pressure difference across the
block (see Eq.~(\ref{eq:pressuredif})), resulting in negatively large
pressures within the bulk of the adhesive and consequently large local
debonding stresses.  The angle $\theta$ is largely close to
$\theta=\pm 90^{\circ}$ for the perpendicular case - i.e. at
$\lambda=1.02$ the director has not rotated much away from its initial
orientation, apart from close to the edges $x\sim\pm0.5$. In the
parallel case there is evident director reorientation, particularly
near the edges of the block $x\sim \pm 0.5$ where the director is at
$\theta=\mp 45^{\circ}$ despite initially being close to $\theta=0$ -
i.e. towards the edges of the block, where the shears are greatest in
magnitude, the director has reoriented to accommodate the shear.

\begin{figure}
\includegraphics[width=0.48\textwidth]{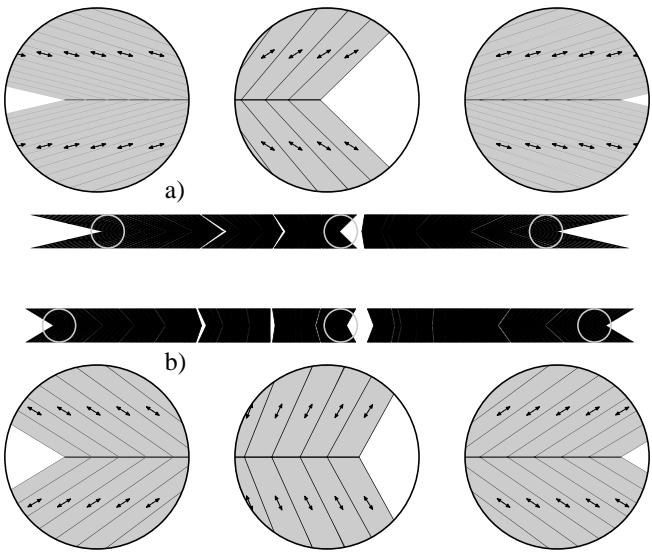}
\caption{A representation of the blocks for the perpendicular (a) and
  parallel (b) alignments.  Some cavities are visible within the
  adhesive layer. The three large circles show close ups on the
  relevant location in the adhesive.  The arrows show the director
  orientation.}
\label{fig:cartoon}
\end{figure}

Fig.~\ref{fig:results}(b) shows how $\sigma_{loc}$, $\lambda_{xy}$ and
$\theta$ vary with position $x$ for $\lambda=1.2$. Comparing with (a)
we can see that $\sigma_{loc}$ has become substantially more negative
for the perpendicular case while the parallel and isotropic results
look broadly similar to the situation in (a).  The shear strains for
the isotropic and parallel situation are similar in magnitude to (a),
but we can now see kinks which correspond to cavities within the
adhesive. The shear strains for the perpendicular case are larger,
both when compared with (a) and when compared with the
parallel/isotropic cases in (b).  The angle $\theta$ is close to $\pm
90^{\circ}$ for the perpendicular case while for the parallel case the
angle remains close to $0^{\circ}$ around $x=0$ but becomes $\pm
50^{\circ}$ towards the edges $x=\pm 0.5$.  Fig.~\ref{fig:cartoon}
shows a representation of the adhesive at $\lambda=1.2$ for the perpendicular (a) and
parallel (b) alignments.
Large cavities are evident. As can be seen, the shears near the edges
of the adhesive are larger for the perpendicular case than for the
parallel case. Note that the traction force on the substrate due to
the adhesive can become negative under some circumstances. When the
shear of the blocks is large, then the constitutive model results in
the expansion of the block and a negative tensile force. The expansion
of rubber under shear is a well known effect \cite{rivlin:444}. It is
interesting to look at the large central cavity for both the
perpendicular and the parallel alignments. We notice in the
perpendicular case that the blocks on either side of the cavity are
sheared in the same sense, while in the parallel case the blocks are
sheared in opposite senses.  This can also be seen in the plot of
$\lambda_{xy}$ in Fig.~\ref{fig:results}(b), the kink in the curves
around $x\sim 0.05$ involves a change in sign for the parallel
alignment, while there is no change in sign for the perpendicular
alignment.

Up until this stage the main contributor to the debonding stress has
been the pressure. As the blocks continue to elongate however we
anticipate the polymer stress contribution will become more and more
important - for a simple neo-Hookean material we would expect these
terms to scale quadratically with $\lambda$.  The rise in the
debonding stress after $\lambda\sim 2$ is principally due to the
polymeric terms in the stress tensor increasing.
Fig.~\ref{fig:results}(c) shows $\sigma_{loc}$, $\lambda_{xy}$ and
$\theta$ as a function of $x$ for $\lambda=3$. We see now that
$\sigma_{loc}$ is somewhat more uniform as a function of position for
all three cases.  The debonding stress is largest for the parallel
alignment and lowest for the perpendicular arrangement.  The debonding
stress is strictly positive for all positions.  It is also interesting
to note that the $x$-range of the plot is substantially reduced at
this strain, indicating that the blocks have slipped.  For both the
parallel and the perpendicular cases we see $\theta\sim 0$ for all
blocks, i.e. the director has largely aligned with the stretch
direction for all blocks.  The shears $\lambda_{xy}$ are smaller in
magnitude than those presented in (b), we notice however they remain
larger in magnitude for the perpendicular alignment than for the
parallel/isotropic.

Above $\lambda \sim 5$ for the isotropic and $\lambda \sim 10$ for the
nematic we see a reduction in the debonding stress.  This is due to
the stress relaxation inherent in the constitutive model of
Eq.~(\ref{eq:scaleddynamicsRR}). We expect the polymer stress to relax
over a time-scale set by $\tau=30$s for the isotropic situation. In
the nematic case two time-scales appear $\tau(1-S)=21s$ and
$\tau(1+2S)=48$s.  The decay time we observe for the perpendicular and
parallel alignments is set by this larger timescale $\tau(1+2S)$.  At
these larger strains the director for each block is aligned along the
$y$-axis and each block has essentially the same value for the local
debonding stress $\sigma_{loc}$.

\begin{figure}[!htb]
  \includegraphics[width=0.48\textwidth]{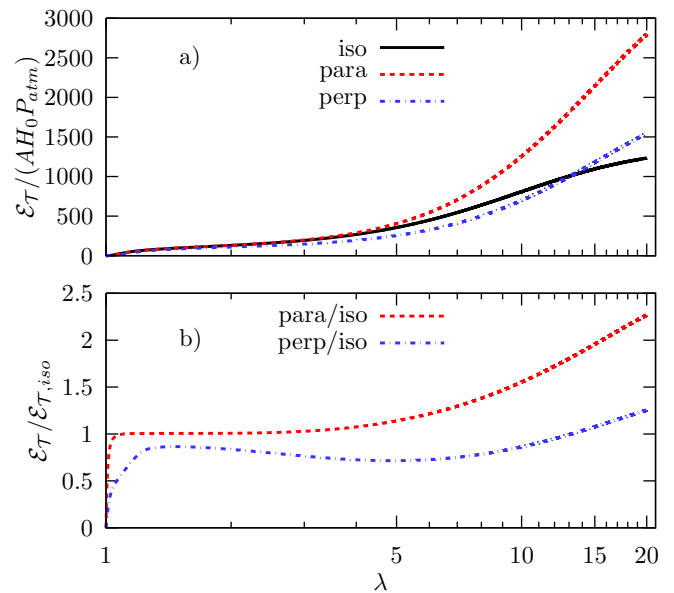}
  \caption{a) The tack energy ${\cal E_{T}}$ as a function of strain
    for the isotropic, parallel and perpendicular cases. b) The
    relative tack of the parallel and perpendicular alignment compared
    to the isotropic case. }
\label{fig:tack2}
\end{figure}
In Fig.~\ref{fig:tack2}(a) the scaled tack energy ${\cal
  E_{T}}/(AH_{0}P_\textrm{atm})$ defined in Eq.~(\ref{eq:tack}) is
plotted as a function of deformation $\lambda$ for the isotropic,
parallel and perpendicular cases. In practice the adhesive eventually
detaches at a particular deformation \cite{glassmaker2008}.  As can be
seen, the tack energy is always largest for the parallel alignment.
The tack energy is mostly lowest for the perpendicular alignment,
although this changes when above $\lambda\sim 13$ where the tack
energy for the isotropic adhesive becomes lower. This plot
demonstrates that the adhesive properties of a nematic adhesive can be
altered both by changing the initial alignment (i.e. switching from
the parallel to the perpendicular geometry) and by changing the order
parameter (i.e. switching between the isotropic and nematic phases).
The latter transition is relatively simple to achieve either by
altering the temperature or by using photo-active
nematics~\cite{corbett:09}, whilst the former might be achievable by
mechanically stretching the adhesive layer. The difference in tack
energy is not so large for small deformations, however it can become
appreciable for larger deformations.  In Fig.~\ref{fig:tack2}(b) we
show the tack energy ratio between the parallel:isotropic and
perpendicular:isotropic as a function of the deformation $\lambda$.
For the parallel:isotropic plot the ratio is close to unity
deformations up to $\lambda\approx 3$, after which we see a gradual
increase in the relative tack up to a value of 2.3 by $\lambda=20$.
The curve for the perpendicular:isotropic ratio is slightly more
complicated.  Initially the relative tack is below unity up until
deformations of around $\lambda\approx 14$, beyond which the relative
tack continues to increase above unity. The detachment process is not
modelled here so we will assume detachment occurs at a strain of
$\lambda = 10$, which is consistent with previous work
\cite{glassmaker2008}. For large strain the precise choice of this
detachment does not change our conclusions. For $\lambda=10$ the ratio
of the tack energies is 1.55:1:0.86 for para:iso:perp.

As can be seen in Table~\ref{table:one} there are a large number of
parameters in our model which we might adjust in order to maximise the
difference in tack between the isotropic and nematic states.  We now
consider how changing several of these parameters changes the force
extension curves and resultant tack.

\subsection{Varying $\tau$}

\begin{figure}[!t]
  \includegraphics[width=0.48\textwidth]{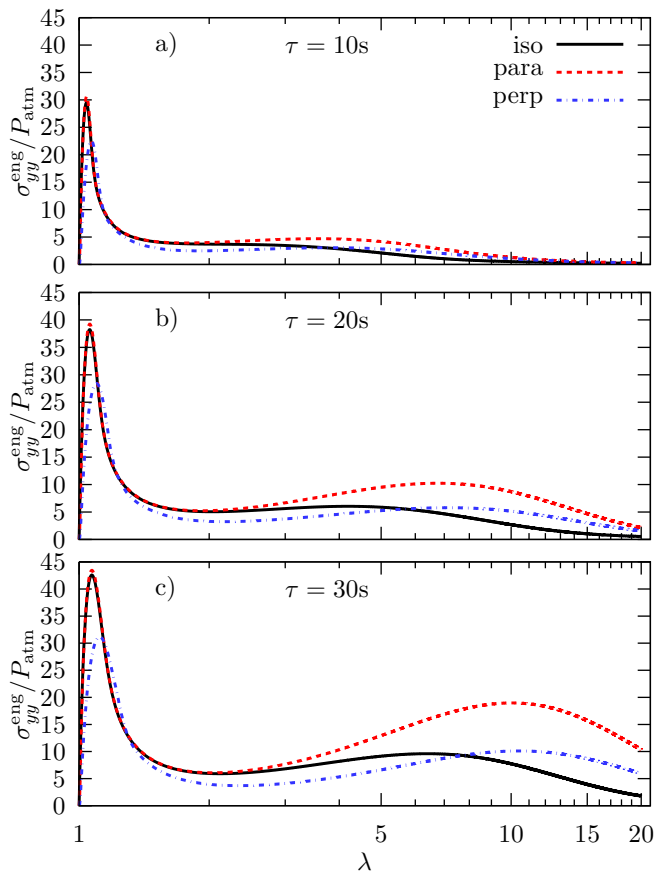}
  \caption{Variation of the debonding stress with increasing deformation
    for several values of $\tau$ shown in the figure. Other parameters
    are as in Table \ref{table:one}. }
\label{fig:tau1}
\end{figure}

The time constant $\tau$ is the fundamental time-scale over which
stress is relaxed away.  In the nematic case we in fact have two
time-scales $\tau_{\perp}=\tau(1-S)$ for the relaxation of stress
perpendicular to the director and $\tau_{\parallel}=\tau(1+2S)$ for
the relaxation of stress parallel to the director. Nevertheless the
average of these time-scales is still $\tau$
($\tau=(\tau_{\parallel}+2\tau_{\perp})/3$).  The time-scale $\tau$ is
also related directly to the viscosity $\eta$ which appears in our
dynamical equation for the cavity radius (see Eq.~(\ref{eq:cavdyn}))
via $\eta=G\tau$. Altering $\tau$ can thus be expected to alter both
the small strain regions of the tack curve where the debonding force
is largely determined by the cavities, and at larger strains where the
debonding force is largely due to the elastic deformation of fibrils.
Fig.~\ref{fig:tau1} shows the force extension curves for three
different values of $\tau$.
All other parameters are as listed in Table~\ref{table:one}.  We can
see that the effect of increasing $\tau$ is to increase the heights of
the two peaks which occur in the plot and to move the peaks to larger
deformations. Therefore a larger $\tau$ tends to produce a greater tack
energy. Of more interest here however perhaps is the relative tack.
For $\lambda=10$ the tack energies are in the ratio 1.57:1:1.11
($\parallel$:iso:$\perp$) for $\tau=10$s, 1.67:1:1 for $\tau=20$s and
1.55:1:0.86 for $\tau=30$s.  The relative tack values are summarsied in Table~\ref{table:two}.

\subsection{Varying $S$}

\begin{figure}[!t]
\includegraphics[width=0.48\textwidth]{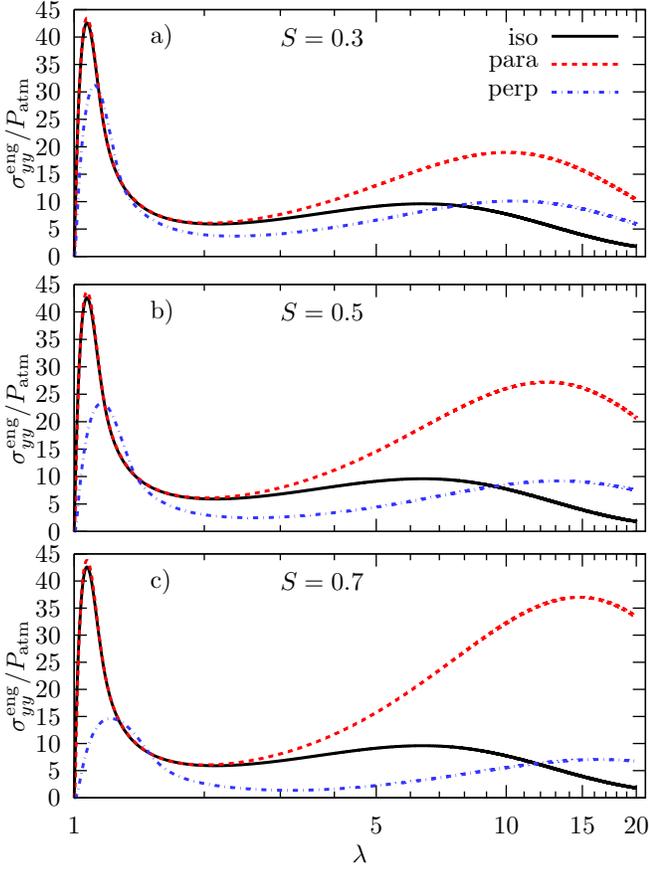}
\caption{Variation of the debonding stress with increasing deformation for
  several values of $S$. Other parameters are as in Table \ref{table:one}.}
\label{fig:S1}
\end{figure}
Figure~\ref{fig:S1} shows the effect of varying the nematic order
parameter $S$ has on the debonding stress as a function of the deformation
$\lambda$.  All other parameters in these plots are as listed in
table~\ref{table:one}. For deformations less than $\lambda\sim 2$ there is
very little difference the curves for the parallel alignment and for
the isotropic adhesive.  The perpendicular alignment is quite strongly
influenced by the order parameter. Increasing the order parameter
leads to a reduction in the height of the first peak in the debonding
stress.  The deformation at which the peak occurs also becomes larger with
increasing order parameter. At larger deformations both the parallel and
perpendicular alignments are effected by changes in the order
parameter.  Increasing the order parameter results in the second peak
in the debonding curves occurring at a larger deformation and a larger
debonding stress, this can be understood by considering the time-scale
for stress relaxation parallel to the director
$\tau_{\parallel}=\tau(1+2S)$.  At larger values of $S$ this
time-scale is longer, so elastic stresses build up for a longer time
for increasing $S$.  For $\lambda=10$ the tack energies are in the ratio
1.55:1:0.86 for $S=0.3$, 1.84:1:0.65 for $S=0.5$ and 2.07:1:0.41 for
$S=0.7$.  Hence there is a reversible change in the tack energy by
more than a factor of $2$ for $S=0.7$ between the parallel and the
isotropic states and between the isotropic and perpendicular states.  The relative tack values are summarsied in Table~\ref{table:two}.

\subsection{Varying $G$}
\begin{figure}[!tb]
\includegraphics[width=0.48\textwidth]{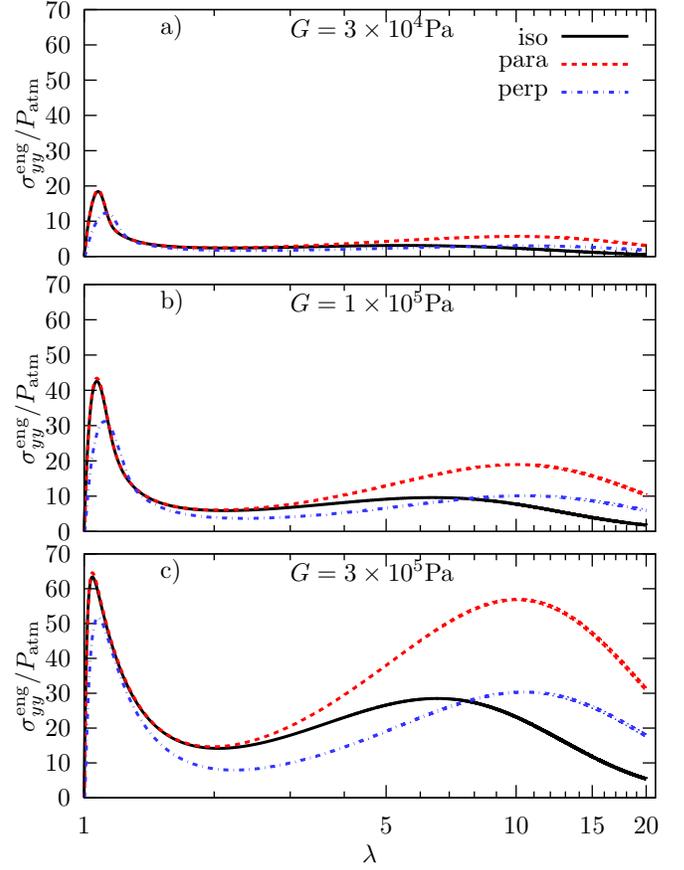}
\caption{Variation of the debonding stress with increasing deformation for
  several values of $G$ shown on the figure. Other parameters
    are as in Table \ref{table:one}.}
\label{fig:G1}
\end{figure}
Figure~\ref{fig:G1} shows the effect that varying the elastic modulus
$G$ has on the debonding stress as a function of the deformation $\lambda$.
All other parameters in these plots are as listed in
Table~\ref{table:one}.  Note the axes have the same scale on all three
plots.  It is apparent that increasing $G$ increases the magnitude of
the debonding stress.  The deformation at which the second peak occurs is
the same in all three plots, the associated debonding stress at the
second peak scales proportionally with $G$.  The debonding stress at
the first peak is larger for larger values of $G$, however the
debonding stress at the first peak is not proportional to $G$ - the
debonding stress at the first peak in fig.~\ref{fig:G1}(c) is roughly
three times greater than in fig.~\ref{fig:G1}(a) while $G$ changes by
a factor of $10$ between the plots.

It is apparent that increasing $G$ leads to greater values for the
absolute tack. The tack values for $\lambda=10$ are in the ratio
1.49:1:0.86 ($\parallel$:iso:$\perp$) for $G=3\times 10^{4}$Pa, 1.55:1:0.86 for
$G=1\times 10^{5}$Pa and 1.61:1:0.86 for $G=3\times 10^{5}$Pa. It is
interesting to note that the relative tack for switching between the
isotropic and the perpendicular states is identical for the three
values of $G$ listed, this is a peculiarity of our choice to quote
results for $\lambda=10$.  The relative tack values are summarsied in Table~\ref{table:two}.

\subsection{Varying $\dot{\lambda}$}

\begin{figure}[!t]
\includegraphics[width=0.48\textwidth]{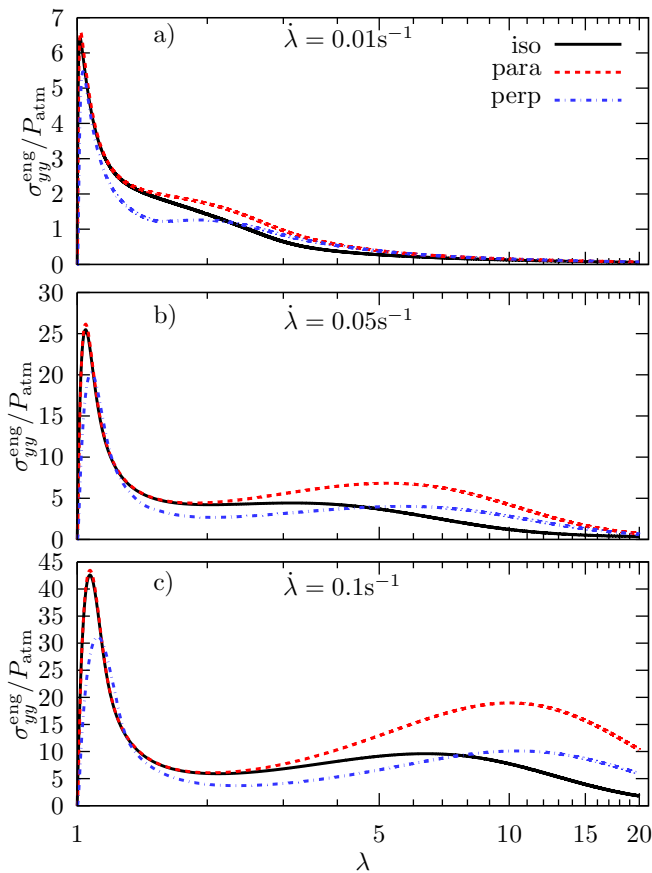}
\caption{Variation of the debonding stress with increasing deformation for
  several values of $\dot{\lambda}$ shown on the figure. Other
  parameters are as in Table \ref{table:one}.}
\label{fig:ldot}
\end{figure}
We have previously considered the effect of changing $\tau$, $S$ and
$G$.  These are material parameters of the adhesive that can be
changed by altering the material chemistry or the chain architecture (chain length, entanglement, molecular weight, chain branching, etc).  We can also consider
changing experimental parameters such as the strain rate
$\dot{\lambda}$.  Figure~\ref{fig:ldot} shows the debonding stress as
a function of the deformation $\lambda$ for several values of
$\dot{\lambda}$.  All other parameters are as listed in
table~\ref{table:one}.  As can be seen, changing $\dot{\lambda}$ has
quite a large effect on the form of the curves.  For
$\dot{\lambda}=0.01$s$^{-1}$ as shown in fig.~\ref{fig:ldot}(a) there
appears to be no second peak in the curves for the parallel and
isotropic adhesives, a second peak is just visible in the curve for
the perpendicular alignment.  This form of the debonding curve is
typical of ``liquid-like'' debonding and is called \emph{cohesive
  debonding}.  In the current case since the Deborah number
$De=\dot{\lambda}\tau = 0.3$ is less than unity the elastic stresses
are able to relax away over the time-scale associated with the
increase in deformation and we are left with a viscous liquid type
debonding curve.  Increasing the strain rate to
$\dot{\lambda}=0.05$s$^{-1}$ produces quite a big difference in the
curves.  The overall magnitude of the stresses has increased, and we
can now clearly see secondary peaks in the debonding stress.  The
shape of this debonding curve is typical of materials intermediate
between a brittle solid and a viscous liquid, and is known as
\emph{adhesive debonding}.  Increasing the strain rate further to
$\dot{\lambda}=0.1$s$^{-1}$ and we see the general form of the curve
is the same as in fig.~\ref{fig:ldot}$(b)$, but the overall magnitude
of the stresses has increased, and the deformation at which the second peak
occurs is larger - the deformation corresponding to a relaxation time $\tau$
is larger for larger strain rates.  These observations are consistent with experimental results on isotropic adhesives~\cite{bellamine:2011}.
It is clear that we obtain larger
absolute tack values for larger strain rates.  Comparing the relative
tack values at $\lambda=10$ we obtain 1.23:1:1.03 for
$\dot{\lambda}=0.01$s$^{-1}$ ($\parallel$:iso:$\perp$), 1.72:1:1.09 for
$\dot{\lambda}=0.05$s$^{-1}$ and 1.55:1:0.86 for
$\dot{\lambda}=0.1$s$^{-1}$.  The relative tack values are summarsied in Table~\ref{table:two}.
\begin{table}[!tb]
\begin{tabular}{|c|c|c|c|}
\hline
&\hspace{2em} &\hspace{2em} $\parallel$:iso\hspace{2em} &\hspace{2em} $\perp$:iso\hspace{2em} \\
\hline
\multirow{3}{*}{$\tau$ (s)} & 10&1.57&1.11\\
&20&1.67&1\\
&30&1.55&0.86\\
\hline
\multirow{3}{*}{$S$} & 0.3&1.55&0.86\\
&0.5&1.84&0.65\\
&0.7&2.07&0.41\\
\hline
\multirow{3}{*}{$G$ ($\times 10^{5}$Pa)} & 0.3&1.49&0.86\\
&1&1.55&0.86\\
&3&1.61&0.86\\
\hline
\multirow{3}{*}{$\dot{\lambda}$ ($s^{-1}$)} & 0.3&1.49&0.86\\
&1&1.55&0.86\\
&3&1.61&0.86\\
\hline
\end{tabular}
\caption{The relative tack values at a deformation of $\lambda=10$ for parallel:isotropic and perpendicular:isotropic for various parameters.}
\label{table:two}
\end{table}
\section{Conclusions}

In this paper we have employed a modified version of the block model
of Yamaguchi \textit{et al.}~\cite{yamaguchimorita2006} and the
nematic dumbbell constitutive equation of Maffetone and
Marucci~\cite{1992maffettone} to model the adhesive debonding of
nematic elastomer adhesives. These models include several
approximations and simplifying assumptions in their description of
adhesive debonding:
\begin{enumerate}
\item The flow of films is assumed to be a superposition of slippage
  and a piecewise linear deformation of the block,

\item The cavities are assumed to be described by gaps between
  adjacent blocks,

\item The cavity dynamics are modelled by the Rayleigh-Plesset equation,

\item The slip velocity at the substrates is assumed to be a linear
  function of the shear stress,

\item The nematic director reorients very quickly compared to the
  polymer relaxation time,

\item The polymers can be described as Gaussian chains,

\item The adhesive cannot debond from the surface or rupture.
\end{enumerate}
Despite these simplifications the main physical processes which occur during debonding are captured and the overall result of our modelling, i.e. demonstrating a reversible difference between the tack energy as a result of the nematic to isotropic transition is valid qualitatively.

Within these modelling assumptions we have shown that there can be
substantial differences in the tack energy when comparing an
isotropic and a nematic adhesive which are due to the difference in
their rheology. For example with a nematic order parameter of $S=0.7$
we were able to achieve a relative tack energy at $\lambda=10$ of
2.07:1 between the parallel aligned nematic and the isotropic and
0.41:1 between the perpendicular aligned nematic and the
isotropic (and thus a ratio of 5.05:1 between the parallel and perpendicular alignments). Given the ease with which one can reversibly cycle between
an isotropic and nematic phase this work gives impetus to the
experimental investigation of this mechanism of switching adhesives on
and off.

\acknowledgements 

We would like to thank Joseph Keddie for helpful discussions. This
work is supported by SEPnet, Wolfson College Oxford, and the EPSRC
through grant EP/I01277X/1.

\bibliography{./LCE_adhesives.bib}

\ifx\mcitethebibliography\mciteundefinedmacro
\PackageError{apsrevM.bst}{mciteplus.sty has not been loaded}
{This bibstyle requires the use of the mciteplus package.}\fi
\begin{mcitethebibliography}{33}
\expandafter\ifx\csname natexlab\endcsname\relax\def\natexlab#1{#1}\fi
\expandafter\ifx\csname bibnamefont\endcsname\relax
  \def\bibnamefont#1{#1}\fi
\expandafter\ifx\csname bibfnamefont\endcsname\relax
  \def\bibfnamefont#1{#1}\fi
\expandafter\ifx\csname citenamefont\endcsname\relax
  \def\citenamefont#1{#1}\fi
\expandafter\ifx\csname url\endcsname\relax
  \def\url#1{\texttt{#1}}\fi
\expandafter\ifx\csname urlprefix\endcsname\relax\def\urlprefix{URL }\fi
\providecommand{\bibinfo}[2]{#2}
\providecommand{\eprint}[2][]{\url{#2}}

\bibitem[{\citenamefont{Gurney et~al.}(2012)\citenamefont{Gurney, Dupin, Nunes,
  Ouzineb, Siband, Asua, Armes, and Keddie}}]{gurney2012}
\bibinfo{author}{\bibfnamefont{R.~S.} \bibnamefont{Gurney}},
  \bibinfo{author}{\bibfnamefont{D.}~\bibnamefont{Dupin}},
  \bibinfo{author}{\bibfnamefont{J.~S.} \bibnamefont{Nunes}},
  \bibinfo{author}{\bibfnamefont{K.}~\bibnamefont{Ouzineb}},
  \bibinfo{author}{\bibfnamefont{E.}~\bibnamefont{Siband}},
  \bibinfo{author}{\bibfnamefont{J.~M.} \bibnamefont{Asua}},
  \bibinfo{author}{\bibfnamefont{S.~P.} \bibnamefont{Armes}}, \bibnamefont{and}
  \bibinfo{author}{\bibfnamefont{J.~L.} \bibnamefont{Keddie}},
  \bibinfo{journal}{Submitted to ACS Applied Materials \& Interfaces}
  (\bibinfo{year}{2012})\relax
\mciteBstWouldAddEndPuncttrue
\mciteSetBstMidEndSepPunct{\mcitedefaultmidpunct}
{\mcitedefaultendpunct}{\mcitedefaultseppunct}\relax
\EndOfBibitem
\bibitem[{\citenamefont{Dahlquist}(1969)}]{Dahlquist1969}
\bibinfo{author}{\bibfnamefont{C.~A.} \bibnamefont{Dahlquist}},
  \emph{\bibinfo{title}{Pressure-Sensitive adhesives}},
  vol.~\bibinfo{volume}{2} of \emph{\bibinfo{series}{Treatise on Adhesion and
  Adhesives}} (\bibinfo{publisher}{Dekker, New York},
  \bibinfo{year}{1969})\relax
\mciteBstWouldAddEndPuncttrue
\mciteSetBstMidEndSepPunct{\mcitedefaultmidpunct}
{\mcitedefaultendpunct}{\mcitedefaultseppunct}\relax
\EndOfBibitem
\bibitem[{\citenamefont{Deplac\'e et~al.}(2009)\citenamefont{Deplac\'e,
  Carelli, Mariot, Retsos, Chateauminois, Ouzineb, and
  Creton}}]{deplacecarelli2009}
\bibinfo{author}{\bibfnamefont{F.}~\bibnamefont{Deplac\'e}},
  \bibinfo{author}{\bibfnamefont{C.}~\bibnamefont{Carelli}},
  \bibinfo{author}{\bibfnamefont{S.}~\bibnamefont{Mariot}},
  \bibinfo{author}{\bibfnamefont{H.}~\bibnamefont{Retsos}},
  \bibinfo{author}{\bibfnamefont{A.}~\bibnamefont{Chateauminois}},
  \bibinfo{author}{\bibfnamefont{K.}~\bibnamefont{Ouzineb}}, \bibnamefont{and}
  \bibinfo{author}{\bibfnamefont{C.}~\bibnamefont{Creton}},
  \bibinfo{journal}{The Journal of Adhesion} \textbf{\bibinfo{volume}{85}},
  \bibinfo{pages}{18} (\bibinfo{year}{2009})\relax
\mciteBstWouldAddEndPuncttrue
\mciteSetBstMidEndSepPunct{\mcitedefaultmidpunct}
{\mcitedefaultendpunct}{\mcitedefaultseppunct}\relax
\EndOfBibitem
\bibitem[{\citenamefont{Lindner et~al.}(2006)\citenamefont{Lindner, Lestriez,
  Mariot, Creton, Maevis, L\"uhmann, and Brummer}}]{lindner2006}
\bibinfo{author}{\bibfnamefont{A.}~\bibnamefont{Lindner}},
  \bibinfo{author}{\bibfnamefont{B.}~\bibnamefont{Lestriez}},
  \bibinfo{author}{\bibfnamefont{S.}~\bibnamefont{Mariot}},
  \bibinfo{author}{\bibfnamefont{C.}~\bibnamefont{Creton}},
  \bibinfo{author}{\bibfnamefont{T.}~\bibnamefont{Maevis}},
  \bibinfo{author}{\bibfnamefont{B.}~\bibnamefont{L\"uhmann}},
  \bibnamefont{and} \bibinfo{author}{\bibfnamefont{R.}~\bibnamefont{Brummer}},
  \bibinfo{journal}{The Journal of Adhesion} \textbf{\bibinfo{volume}{82}},
  \bibinfo{pages}{267} (\bibinfo{year}{2006})\relax
\mciteBstWouldAddEndPuncttrue
\mciteSetBstMidEndSepPunct{\mcitedefaultmidpunct}
{\mcitedefaultendpunct}{\mcitedefaultseppunct}\relax
\EndOfBibitem
\bibitem[{\citenamefont{Deplace et~al.}(2009)\citenamefont{Deplace, Rabjohns,
  Yamaguchi, Foster, Carelli, Lei, Ouzineb, Keddie, Lovell, and
  Creton}}]{deplacerabjohns2009}
\bibinfo{author}{\bibfnamefont{F.}~\bibnamefont{Deplace}},
  \bibinfo{author}{\bibfnamefont{M.~A.} \bibnamefont{Rabjohns}},
  \bibinfo{author}{\bibfnamefont{T.}~\bibnamefont{Yamaguchi}},
  \bibinfo{author}{\bibfnamefont{A.~B.} \bibnamefont{Foster}},
  \bibinfo{author}{\bibfnamefont{C.}~\bibnamefont{Carelli}},
  \bibinfo{author}{\bibfnamefont{C.-H.} \bibnamefont{Lei}},
  \bibinfo{author}{\bibfnamefont{K.}~\bibnamefont{Ouzineb}},
  \bibinfo{author}{\bibfnamefont{J.~L.} \bibnamefont{Keddie}},
  \bibinfo{author}{\bibfnamefont{P.~A.} \bibnamefont{Lovell}},
  \bibnamefont{and} \bibinfo{author}{\bibfnamefont{C.}~\bibnamefont{Creton}},
  \bibinfo{journal}{Soft Matter} \textbf{\bibinfo{volume}{5}},
  \bibinfo{pages}{1440} (\bibinfo{year}{2009})\relax
\mciteBstWouldAddEndPuncttrue
\mciteSetBstMidEndSepPunct{\mcitedefaultmidpunct}
{\mcitedefaultendpunct}{\mcitedefaultseppunct}\relax
\EndOfBibitem
\bibitem[{\citenamefont{Gay and Leibler}(1999)}]{gay1999}
\bibinfo{author}{\bibfnamefont{C.}~\bibnamefont{Gay}} \bibnamefont{and}
  \bibinfo{author}{\bibfnamefont{L.}~\bibnamefont{Leibler}},
  \bibinfo{journal}{Phys. Rev. Lett.} \textbf{\bibinfo{volume}{82}},
  \bibinfo{pages}{936} (\bibinfo{year}{1999})\relax
\mciteBstWouldAddEndPuncttrue
\mciteSetBstMidEndSepPunct{\mcitedefaultmidpunct}
{\mcitedefaultendpunct}{\mcitedefaultseppunct}\relax
\EndOfBibitem
\bibitem[{\citenamefont{Yamaguchi et~al.}(2006)\citenamefont{Yamaguchi, Morita,
  and Doi}}]{yamaguchimorita2006}
\bibinfo{author}{\bibfnamefont{T.}~\bibnamefont{Yamaguchi}},
  \bibinfo{author}{\bibfnamefont{H.}~\bibnamefont{Morita}}, \bibnamefont{and}
  \bibinfo{author}{\bibfnamefont{M.}~\bibnamefont{Doi}}, \bibinfo{journal}{Eur.
  Phys. J. E} \textbf{\bibinfo{volume}{20}}, \bibinfo{pages}{7}
  (\bibinfo{year}{2006})\relax
\mciteBstWouldAddEndPuncttrue
\mciteSetBstMidEndSepPunct{\mcitedefaultmidpunct}
{\mcitedefaultendpunct}{\mcitedefaultseppunct}\relax
\EndOfBibitem
\bibitem[{\citenamefont{Yamaguchi and Doi}(2006)}]{yamaguchidoi2006}
\bibinfo{author}{\bibfnamefont{T.}~\bibnamefont{Yamaguchi}} \bibnamefont{and}
  \bibinfo{author}{\bibfnamefont{M.}~\bibnamefont{Doi}}, \bibinfo{journal}{Eur.
  Phys. J. E} \textbf{\bibinfo{volume}{21}}, \bibinfo{pages}{331}
  (\bibinfo{year}{2006})\relax
\mciteBstWouldAddEndPuncttrue
\mciteSetBstMidEndSepPunct{\mcitedefaultmidpunct}
{\mcitedefaultendpunct}{\mcitedefaultseppunct}\relax
\EndOfBibitem
\bibitem[{\citenamefont{Webster}(1999)}]{webster1999}
\bibinfo{author}{\bibfnamefont{I.}~\bibnamefont{Webster}},
  \bibinfo{journal}{Int. J. adhesion adhesives} \textbf{\bibinfo{volume}{19}},
  \bibinfo{pages}{29} (\bibinfo{year}{1999})\relax
\mciteBstWouldAddEndPuncttrue
\mciteSetBstMidEndSepPunct{\mcitedefaultmidpunct}
{\mcitedefaultendpunct}{\mcitedefaultseppunct}\relax
\EndOfBibitem
\bibitem[{\citenamefont{Boyne et~al.}(2001)\citenamefont{Boyne, Millan, and
  Webster}}]{Boyne:2001}
\bibinfo{author}{\bibfnamefont{J.~M.} \bibnamefont{Boyne}},
  \bibinfo{author}{\bibfnamefont{E.~J.} \bibnamefont{Millan}},
  \bibnamefont{and} \bibinfo{author}{\bibfnamefont{I.}~\bibnamefont{Webster}},
  \bibinfo{journal}{International Journal of Adhesion and Adhesives}
  \textbf{\bibinfo{volume}{21}}, \bibinfo{pages}{49}
  (\bibinfo{year}{2001})\relax
\mciteBstWouldAddEndPuncttrue
\mciteSetBstMidEndSepPunct{\mcitedefaultmidpunct}
{\mcitedefaultendpunct}{\mcitedefaultseppunct}\relax
\EndOfBibitem
\bibitem[{\citenamefont{Trenor et~al.}(2005)\citenamefont{Trenor, Long, and
  J.}}]{Trenor:2005}
\bibinfo{author}{\bibfnamefont{S.~J.} \bibnamefont{Trenor}},
  \bibinfo{author}{\bibfnamefont{T.~E.} \bibnamefont{Long}}, \bibnamefont{and}
  \bibinfo{author}{\bibfnamefont{L.~B.} \bibnamefont{J.}},
  \bibinfo{journal}{The Journal of Adhesion} \textbf{\bibinfo{volume}{81}},
  \bibinfo{pages}{213} (\bibinfo{year}{2005})\relax
\mciteBstWouldAddEndPuncttrue
\mciteSetBstMidEndSepPunct{\mcitedefaultmidpunct}
{\mcitedefaultendpunct}{\mcitedefaultseppunct}\relax
\EndOfBibitem
\bibitem[{\citenamefont{Diethert et~al.}(2011)\citenamefont{Diethert, Ecker,
  Peykova, Willenbacher, and M\"{u}ller-Buschbaum}}]{Diethert:2011}
\bibinfo{author}{\bibfnamefont{A.}~\bibnamefont{Diethert}},
  \bibinfo{author}{\bibfnamefont{K.}~\bibnamefont{Ecker}},
  \bibinfo{author}{\bibfnamefont{Y.}~\bibnamefont{Peykova}},
  \bibinfo{author}{\bibfnamefont{N.}~\bibnamefont{Willenbacher}},
  \bibnamefont{and}
  \bibinfo{author}{\bibfnamefont{P.}~\bibnamefont{M\"{u}ller-Buschbaum}},
  \bibinfo{journal}{ACS Applied Materials \& Interfaces}
  \textbf{\bibinfo{volume}{3}}, \bibinfo{pages}{2012}
  (\bibinfo{year}{2011})\relax
\mciteBstWouldAddEndPuncttrue
\mciteSetBstMidEndSepPunct{\mcitedefaultmidpunct}
{\mcitedefaultendpunct}{\mcitedefaultseppunct}\relax
\EndOfBibitem
\bibitem[{\citenamefont{Stuart et~al.}(2010)\citenamefont{Stuart, Huck, Genzer,
  M\"{u}ller, Ober, Stamm, Sukhorukov, Szleifer, Tsukruk, and
  Urban}}]{Stuart:2010}
\bibinfo{author}{\bibfnamefont{M.~A.~C.} \bibnamefont{Stuart}},
  \bibinfo{author}{\bibfnamefont{W.~T.~S.} \bibnamefont{Huck}},
  \bibinfo{author}{\bibfnamefont{J.}~\bibnamefont{Genzer}},
  \bibinfo{author}{\bibfnamefont{M.}~\bibnamefont{M\"{u}ller}},
  \bibinfo{author}{\bibfnamefont{C.}~\bibnamefont{Ober}},
  \bibinfo{author}{\bibfnamefont{M.}~\bibnamefont{Stamm}},
  \bibinfo{author}{\bibfnamefont{G.~B.} \bibnamefont{Sukhorukov}},
  \bibinfo{author}{\bibfnamefont{I.}~\bibnamefont{Szleifer}},
  \bibinfo{author}{\bibfnamefont{V.~V.} \bibnamefont{Tsukruk}},
  \bibnamefont{and} \bibinfo{author}{\bibfnamefont{M.~W.} \bibnamefont{Urban}},
  \bibinfo{journal}{Nature Materials} \textbf{\bibinfo{volume}{9}},
  \bibinfo{pages}{101} (\bibinfo{year}{2010})\relax
\mciteBstWouldAddEndPuncttrue
\mciteSetBstMidEndSepPunct{\mcitedefaultmidpunct}
{\mcitedefaultendpunct}{\mcitedefaultseppunct}\relax
\EndOfBibitem
\bibitem[{\citenamefont{La~Spina et~al.}(2007)\citenamefont{La~Spina,
  Tomlinson, Ruiz-P\'{e}rez, Chiche, Langridge, and Geoghegan}}]{Spina:2007}
\bibinfo{author}{\bibfnamefont{R.}~\bibnamefont{La~Spina}},
  \bibinfo{author}{\bibfnamefont{M.~R.} \bibnamefont{Tomlinson}},
  \bibinfo{author}{\bibfnamefont{L.}~\bibnamefont{Ruiz-P\'{e}rez}},
  \bibinfo{author}{\bibfnamefont{A.}~\bibnamefont{Chiche}},
  \bibinfo{author}{\bibfnamefont{S.}~\bibnamefont{Langridge}},
  \bibnamefont{and}
  \bibinfo{author}{\bibfnamefont{M.}~\bibnamefont{Geoghegan}},
  \bibinfo{journal}{Angewandte Chemie International Edition}
  \textbf{\bibinfo{volume}{46}}, \bibinfo{pages}{6460}
  (\bibinfo{year}{2007})\relax
\mciteBstWouldAddEndPuncttrue
\mciteSetBstMidEndSepPunct{\mcitedefaultmidpunct}
{\mcitedefaultendpunct}{\mcitedefaultseppunct}\relax
\EndOfBibitem
\bibitem[{\citenamefont{de~Crevoisier et~al.}(1999)\citenamefont{de~Crevoisier,
  Fabre, Corpart, and Leibler}}]{crevoisier1999}
\bibinfo{author}{\bibfnamefont{G.}~\bibnamefont{de~Crevoisier}},
  \bibinfo{author}{\bibfnamefont{P.}~\bibnamefont{Fabre}},
  \bibinfo{author}{\bibfnamefont{J.-M.} \bibnamefont{Corpart}},
  \bibnamefont{and} \bibinfo{author}{\bibfnamefont{L.}~\bibnamefont{Leibler}},
  \bibinfo{journal}{Science} \textbf{\bibinfo{volume}{285}},
  \bibinfo{pages}{1246} (\bibinfo{year}{1999})\relax
\mciteBstWouldAddEndPuncttrue
\mciteSetBstMidEndSepPunct{\mcitedefaultmidpunct}
{\mcitedefaultendpunct}{\mcitedefaultseppunct}\relax
\EndOfBibitem
\bibitem[{\citenamefont{Kamperman and Synytska}(2012)}]{Kamperman:2012}
\bibinfo{author}{\bibfnamefont{M.}~\bibnamefont{Kamperman}} \bibnamefont{and}
  \bibinfo{author}{\bibfnamefont{A.}~\bibnamefont{Synytska}},
  \bibinfo{journal}{J. Mater. Chem.} pp.~\bibinfo{pages}{--}
  (\bibinfo{year}{2012}),
  \urlprefix\url{http://dx.doi.org/10.1039/C2JM31747H}\relax
\mciteBstWouldAddEndPuncttrue
\mciteSetBstMidEndSepPunct{\mcitedefaultmidpunct}
{\mcitedefaultendpunct}{\mcitedefaultseppunct}\relax
\EndOfBibitem
\bibitem[{\citenamefont{Zubarev et~al.}(1996)\citenamefont{Zubarev, Talroze,
  Yuranova, Vasilets, and Plate}}]{zubarev1996}
\bibinfo{author}{\bibfnamefont{E.~R.} \bibnamefont{Zubarev}},
  \bibinfo{author}{\bibfnamefont{R.~V.} \bibnamefont{Talroze}},
  \bibinfo{author}{\bibfnamefont{T.~I.} \bibnamefont{Yuranova}},
  \bibinfo{author}{\bibfnamefont{V.~N.} \bibnamefont{Vasilets}},
  \bibnamefont{and} \bibinfo{author}{\bibfnamefont{N.~A.} \bibnamefont{Plate}},
  \bibinfo{journal}{Macromol. Rapid Commun.} \textbf{\bibinfo{volume}{17}},
  \bibinfo{pages}{43} (\bibinfo{year}{1996})\relax
\mciteBstWouldAddEndPuncttrue
\mciteSetBstMidEndSepPunct{\mcitedefaultmidpunct}
{\mcitedefaultendpunct}{\mcitedefaultseppunct}\relax
\EndOfBibitem
\bibitem[{\citenamefont{K\"upfer and Finkelmann}(1994)}]{kupfer1994}
\bibinfo{author}{\bibfnamefont{J.}~\bibnamefont{K\"upfer}} \bibnamefont{and}
  \bibinfo{author}{\bibfnamefont{H.}~\bibnamefont{Finkelmann}},
  \bibinfo{journal}{Macromol. Chem. Phys.} \textbf{\bibinfo{volume}{195}},
  \bibinfo{pages}{1353} (\bibinfo{year}{1994})\relax
\mciteBstWouldAddEndPuncttrue
\mciteSetBstMidEndSepPunct{\mcitedefaultmidpunct}
{\mcitedefaultendpunct}{\mcitedefaultseppunct}\relax
\EndOfBibitem
\bibitem[{\citenamefont{Warner and Terentjev}(2007)}]{warnerterentjev2007}
\bibinfo{author}{\bibfnamefont{M.}~\bibnamefont{Warner}} \bibnamefont{and}
  \bibinfo{author}{\bibfnamefont{E.~M.} \bibnamefont{Terentjev}},
  \emph{\bibinfo{title}{Liquid Crystal Elastomers}} (\bibinfo{publisher}{Oxford
  University Press, Oxford}, \bibinfo{year}{2007})\relax
\mciteBstWouldAddEndPuncttrue
\mciteSetBstMidEndSepPunct{\mcitedefaultmidpunct}
{\mcitedefaultendpunct}{\mcitedefaultseppunct}\relax
\EndOfBibitem
\bibitem[{\citenamefont{Olmsted}(1994)}]{Olmsted1994}
\bibinfo{author}{\bibfnamefont{P.~D.} \bibnamefont{Olmsted}},
  \bibinfo{journal}{J. Phys. II France} \textbf{\bibinfo{volume}{4}},
  \bibinfo{pages}{2215} (\bibinfo{year}{1994})\relax
\mciteBstWouldAddEndPuncttrue
\mciteSetBstMidEndSepPunct{\mcitedefaultmidpunct}
{\mcitedefaultendpunct}{\mcitedefaultseppunct}\relax
\EndOfBibitem
\bibitem[{\citenamefont{Maffettone and Marrucci}(1992)}]{1992maffettone}
\bibinfo{author}{\bibfnamefont{P.~L.} \bibnamefont{Maffettone}}
  \bibnamefont{and} \bibinfo{author}{\bibfnamefont{G.}~\bibnamefont{Marrucci}},
  \bibinfo{journal}{J. Rheol.} \textbf{\bibinfo{volume}{36}},
  \bibinfo{pages}{154} (\bibinfo{year}{1992})\relax
\mciteBstWouldAddEndPuncttrue
\mciteSetBstMidEndSepPunct{\mcitedefaultmidpunct}
{\mcitedefaultendpunct}{\mcitedefaultseppunct}\relax
\EndOfBibitem
\bibitem[{\citenamefont{Rayleigh}(1917)}]{rayleigh1917}
\bibinfo{author}{\bibfnamefont{J.~S.} \bibnamefont{Rayleigh}},
  \bibinfo{journal}{Phil. Mag.} \textbf{\bibinfo{volume}{34}},
  \bibinfo{pages}{94} (\bibinfo{year}{1917})\relax
\mciteBstWouldAddEndPuncttrue
\mciteSetBstMidEndSepPunct{\mcitedefaultmidpunct}
{\mcitedefaultendpunct}{\mcitedefaultseppunct}\relax
\EndOfBibitem
\bibitem[{\citenamefont{Plesset and Prosperetti}(1977)}]{plesset1977}
\bibinfo{author}{\bibfnamefont{M.~S.} \bibnamefont{Plesset}} \bibnamefont{and}
  \bibinfo{author}{\bibfnamefont{A.}~\bibnamefont{Prosperetti}},
  \bibinfo{journal}{Ann. Rev. Fluid Mech.} \textbf{\bibinfo{volume}{9}},
  \bibinfo{pages}{145} (\bibinfo{year}{1977})\relax
\mciteBstWouldAddEndPuncttrue
\mciteSetBstMidEndSepPunct{\mcitedefaultmidpunct}
{\mcitedefaultendpunct}{\mcitedefaultseppunct}\relax
\EndOfBibitem
\bibitem[{\citenamefont{Lubensky et~al.}(2002)\citenamefont{Lubensky,
  Mukhopadhyay, Radzihovsky, and Xing}}]{PhysRevE.66.011702}
\bibinfo{author}{\bibfnamefont{T.~C.} \bibnamefont{Lubensky}},
  \bibinfo{author}{\bibfnamefont{R.}~\bibnamefont{Mukhopadhyay}},
  \bibinfo{author}{\bibfnamefont{L.}~\bibnamefont{Radzihovsky}},
  \bibnamefont{and} \bibinfo{author}{\bibfnamefont{X.}~\bibnamefont{Xing}},
  \bibinfo{journal}{Phys. Rev. E} \textbf{\bibinfo{volume}{66}},
  \bibinfo{pages}{011702} (\bibinfo{year}{2002})\relax
\mciteBstWouldAddEndPuncttrue
\mciteSetBstMidEndSepPunct{\mcitedefaultmidpunct}
{\mcitedefaultendpunct}{\mcitedefaultseppunct}\relax
\EndOfBibitem
\bibitem[{\citenamefont{Golubovi\ifmmode~\acute{c}\else \'{c}\fi{} and
  Lubensky}(1989)}]{PhysRevLett.63.1082}
\bibinfo{author}{\bibfnamefont{L.}~\bibnamefont{Golubovi\ifmmode~\acute{c}\else
  \'{c}\fi{}}} \bibnamefont{and} \bibinfo{author}{\bibfnamefont{T.~C.}
  \bibnamefont{Lubensky}}, \bibinfo{journal}{Phys. Rev. Lett.}
  \textbf{\bibinfo{volume}{63}}, \bibinfo{pages}{1082}
  (\bibinfo{year}{1989})\relax
\mciteBstWouldAddEndPuncttrue
\mciteSetBstMidEndSepPunct{\mcitedefaultmidpunct}
{\mcitedefaultendpunct}{\mcitedefaultseppunct}\relax
\EndOfBibitem
\bibitem[{\citenamefont{Chiche et~al.}(2005)\citenamefont{Chiche, Dollhofer,
  and Creton}}]{ISI:000232026600001}
\bibinfo{author}{\bibfnamefont{A.}~\bibnamefont{Chiche}},
  \bibinfo{author}{\bibfnamefont{J.}~\bibnamefont{Dollhofer}},
  \bibnamefont{and} \bibinfo{author}{\bibfnamefont{C.}~\bibnamefont{Creton}},
  \bibinfo{journal}{Eur. Phys. J. E} \textbf{\bibinfo{volume}{17}},
  \bibinfo{pages}{389} (\bibinfo{year}{2005})\relax
\mciteBstWouldAddEndPuncttrue
\mciteSetBstMidEndSepPunct{\mcitedefaultmidpunct}
{\mcitedefaultendpunct}{\mcitedefaultseppunct}\relax
\EndOfBibitem
\bibitem[{\citenamefont{Glassmaker et~al.}(2008)\citenamefont{Glassmaker, Hui,
  Yamaguchi, and Creton}}]{glassmaker2008}
\bibinfo{author}{\bibfnamefont{N.~J.} \bibnamefont{Glassmaker}},
  \bibinfo{author}{\bibfnamefont{C.~Y.} \bibnamefont{Hui}},
  \bibinfo{author}{\bibfnamefont{T.}~\bibnamefont{Yamaguchi}},
  \bibnamefont{and} \bibinfo{author}{\bibfnamefont{C.}~\bibnamefont{Creton}},
  \bibinfo{journal}{Eur. Phys. J. E} \textbf{\bibinfo{volume}{25}},
  \bibinfo{pages}{253} (\bibinfo{year}{2008})\relax
\mciteBstWouldAddEndPuncttrue
\mciteSetBstMidEndSepPunct{\mcitedefaultmidpunct}
{\mcitedefaultendpunct}{\mcitedefaultseppunct}\relax
\EndOfBibitem
\bibitem[{\citenamefont{Maier and Saupe}(1959)}]{Maier:59}
\bibinfo{author}{\bibfnamefont{W.}~\bibnamefont{Maier}} \bibnamefont{and}
  \bibinfo{author}{\bibfnamefont{A.}~\bibnamefont{Saupe}}, \bibinfo{journal}{Z.
  Naturforsch} \textbf{\bibinfo{volume}{14a}}, \bibinfo{pages}{882}
  (\bibinfo{year}{1959})\relax
\mciteBstWouldAddEndPuncttrue
\mciteSetBstMidEndSepPunct{\mcitedefaultmidpunct}
{\mcitedefaultendpunct}{\mcitedefaultseppunct}\relax
\EndOfBibitem
\bibitem[{\citenamefont{Tajbakhsh and Terentjev}(2001)}]{Tajbakhsh:01}
\bibinfo{author}{\bibfnamefont{A.~R.} \bibnamefont{Tajbakhsh}}
  \bibnamefont{and} \bibinfo{author}{\bibfnamefont{E.~M.}
  \bibnamefont{Terentjev}}, \bibinfo{journal}{Eur. Phys. J. E}
  \textbf{\bibinfo{volume}{6}}, \bibinfo{pages}{181}
  (\bibinfo{year}{2001})\relax
\mciteBstWouldAddEndPuncttrue
\mciteSetBstMidEndSepPunct{\mcitedefaultmidpunct}
{\mcitedefaultendpunct}{\mcitedefaultseppunct}\relax
\EndOfBibitem
\bibitem[{\citenamefont{Degrandi}(2009)}]{Degrandi:09}
\bibinfo{author}{\bibfnamefont{M.~E.} \bibnamefont{Degrandi}},
  \emph{\bibinfo{title}{PhD Thesis: Latex Hybrides Urethane/Acrylique pour
  Applications Adhesives}} (\bibinfo{publisher}{Universit\'e Pierre et Marie
  Curie}, \bibinfo{year}{2009})\relax
\mciteBstWouldAddEndPuncttrue
\mciteSetBstMidEndSepPunct{\mcitedefaultmidpunct}
{\mcitedefaultendpunct}{\mcitedefaultseppunct}\relax
\EndOfBibitem
\bibitem[{\citenamefont{Rivlin}(1947)}]{rivlin:444}
\bibinfo{author}{\bibfnamefont{R.~S.} \bibnamefont{Rivlin}},
  \bibinfo{journal}{Journal of Applied Physics} \textbf{\bibinfo{volume}{18}},
  \bibinfo{pages}{444} (\bibinfo{year}{1947})\relax
\mciteBstWouldAddEndPuncttrue
\mciteSetBstMidEndSepPunct{\mcitedefaultmidpunct}
{\mcitedefaultendpunct}{\mcitedefaultseppunct}\relax
\EndOfBibitem
\bibitem[{\citenamefont{Corbett and Warner}(2009)}]{corbett:09}
\bibinfo{author}{\bibfnamefont{D.}~\bibnamefont{Corbett}} \bibnamefont{and}
  \bibinfo{author}{\bibfnamefont{M.}~\bibnamefont{Warner}},
  \bibinfo{journal}{Liquid Crystals} \textbf{\bibinfo{volume}{36}},
  \bibinfo{pages}{1263} (\bibinfo{year}{2009})\relax
\mciteBstWouldAddEndPuncttrue
\mciteSetBstMidEndSepPunct{\mcitedefaultmidpunct}
{\mcitedefaultendpunct}{\mcitedefaultseppunct}\relax
\EndOfBibitem
\bibitem[{\citenamefont{Bellamine et~al.}(2011)\citenamefont{Bellamine,
  Degrandi, Gerst, Stark, Beyers, and Creton}}]{bellamine:2011}
\bibinfo{author}{\bibfnamefont{A.}~\bibnamefont{Bellamine}},
  \bibinfo{author}{\bibfnamefont{E.}~\bibnamefont{Degrandi}},
  \bibinfo{author}{\bibfnamefont{M.}~\bibnamefont{Gerst}},
  \bibinfo{author}{\bibfnamefont{R.}~\bibnamefont{Stark}},
  \bibinfo{author}{\bibfnamefont{C.}~\bibnamefont{Beyers}}, \bibnamefont{and}
  \bibinfo{author}{\bibfnamefont{C.}~\bibnamefont{Creton}},
  \bibinfo{journal}{Macromolecular Materials and Engineering}
  \textbf{\bibinfo{volume}{296}}, \bibinfo{pages}{31} (\bibinfo{year}{2011}),
  ISSN \bibinfo{issn}{1439-2054},
  \urlprefix\url{http://dx.doi.org/10.1002/mame.201000265}\relax
\mciteBstWouldAddEndPuncttrue
\mciteSetBstMidEndSepPunct{\mcitedefaultmidpunct}
{\mcitedefaultendpunct}{\mcitedefaultseppunct}\relax
\EndOfBibitem
\end{mcitethebibliography}
\end{document}